\documentclass[lettersize,journal]{IEEEtran}
\usepackage{amsmath,amsfonts}
\usepackage{algorithmic}
\usepackage{algorithm}
\usepackage{array}
\usepackage[caption=false,font=normalsize,labelfont=sf,textfont=sf]{subfig}
\usepackage{textcomp}
\usepackage{stfloats}
\usepackage{url}
\usepackage{verbatim}
\usepackage{graphicx}
\usepackage{cite}
\usepackage{bm}
\usepackage{siunitx}
\usepackage{comment}
\usepackage{caption}
\usepackage{tikz}
\renewcommand\fbox{\fcolorbox{red}{white}}
\newcommand\submittedtext{%
  \footnotesize This work has been submitted to the IEEE for possible publication. Copyright may be transferred without notice, after which this version may no longer be accessible.}

\newcommand\submittednotice{%
\begin{tikzpicture}[remember picture,overlay]
\node[anchor=south,yshift=10pt] at (current page.south) {\fbox{\parbox{\dimexpr0.65\textwidth-\fboxsep-\fboxrule\relax}{\submittedtext}}};
\end{tikzpicture}%
}
\newcommand\copyrighttext{%
  \footnotesize \textcopyright \the\year{} IEEE. Personal use of this material is permitted. Permission from IEEE must be obtained for all other uses, including reprinting/republishing this material for advertising or promotional purposes, collecting new works for resale or redistribution to servers or lists, or reuse of any copyrighted component of this work in other works.}

\begin{document}

\title{SCE-NTT: A Hardware Accelerator for Number Theoretic Transform Using Superconductor Electronics}

\author{S. Razmkhah, M. Li, Z. Cheng, R. S. Aviles, K. Jackman,~\IEEEmembership{IEEE Member,} J. Delport,~\IEEEmembership{IEEE Member,} L. Schindler,~\IEEEmembership{IEEE Member,} W. Luo, T. Suzuki, M. Kamal,~\IEEEmembership{IEEE Senior Member,} C. L. Ayala,~\IEEEmembership{IEEE Senior Member,} C. J. Fourie,~\IEEEmembership{IEEE Senior Member,} N. Yoshikawa,~\IEEEmembership{IEEE Fellow,} P. A. Beerel, ~\IEEEmembership{IEEE Senior Member,}  S. Gupta, M. Pedram,~\IEEEmembership{IEEE Fellow}
\thanks{S~Razmkhah and M~Li contributed equally to this work and are considered co-first authors.}
\thanks{S~Razmkhah, M~Li, Z~Cheng, R~Aviles, M~Kamal, P~Beerel, S~Gupta, and M~Pedram are with the Ming Hsieh Department of Electrical Engineering, University of Southern California, Los Angeles, California, USA (e-mail: razmkhah@usc.edu, mingyel@usc.edu, zemingch@usc.edu, rsaviles@usc.edu, mehdi.kamal@usc.edu, pabeerel@usc.edu, sandeep@usc.edu, pedram@usc.edu)}
\thanks{CJ Fourie, L Schindler, K Jackman, J Delport are with Stellenbosch University, Stellenbosch, South Africa (e-mail: coenrad@sun.ac.za, lschindler@sun.ac.za, kjackman@sun.ac.za, jdelport@sun.ac.za)}
\thanks{CL Ayala, W. Luo, T. Suzuki, and N Yoshikawa are with the Institute of Advanced Sciences, Yokohama National University, Yokohama, Japan (e-mail: ayala-christopher-pz@ynu.ac.jp, luo-wenhui-dv@ynu.jp, suzuki-takuya-dg@ynu.jp, yoshikawa-nobuyuki-gt@ynu.ac.jp)}}

\markboth{ }%
{Shell \MakeLowercase{\textit{et al.}}: A Sample Article Using IEEEtran.cls for IEEE Journals}


\maketitle

\submittednotice

\begin{abstract}
This research explores the use of superconductor electronics (SCE) for accelerating fully homomorphic encryption (FHE), focusing on the Number-Theoretic Transform (NTT)—a key computational bottleneck in FHE schemes. We present \textbf{SCE-NTT}, a dedicated hardware accelerator based on superconductive single flux quantum (SFQ) logic and memory, targeting high performance and energy efficiency beyond the limits of conventional CMOS. To address SFQ constraints such as limited dense RAM and restricted fanin/fanout, we propose a deeply pipelined NTT-128 architecture using shift register memory (SRM). Designed for $N = 128$ 32-bit coefficients, NTT-128 comprises $\log_2(N) = 7$ processing elements (PEs), each featuring a butterfly unit (BU), dual coefficient memories operating in ping-pong mode via FIFO-based SRM queues, and twiddle factor buffers. The BU integrates a Shoup modular multiplier optimized for a small area, leveraging precomputed twiddle factors. A new RSFQ cell library with over 50 parameterized cells, including compound logic units, was developed for implementation. Functional and timing correctness were validated using JoSIM analog simulations and Verilog models. A multiphase clocking scheme was employed to enhance robustness and reduce path-balancing overhead, improving circuit reliability. Fabricated results show the NTT-128 unit achieves 531 million NTT/sec at 34 GHz, over 100$\times$ faster than state-of-the-art CMOS equivalents. We also project that the architecture can scale to larger sizes, such as a $2^{14}$-point NTT in approximately 482 ns. Key-switch throughput is estimated at 1.63 million operations/sec, significantly exceeding existing hardware. These results demonstrate the strong potential of SCE-based accelerators for scalable, energy-efficient secure computation in the post-quantum era, with further gains anticipated through advances in fabrication.

\end{abstract}

\begin{IEEEkeywords}
Fully homomorphic encryption, number-theoretical transformation, superconductor electronics, shift register memory
\end{IEEEkeywords}

\section{Introduction}
Superconductor electronics (SCE) is a promising technology for computation and accelerators in the post-CMOS era \cite{noauthor_irds_nodate}. Despite the current lack of advanced fabrication processes for SCEs, the high frequency and non-dissipative nature of these circuits provides a two to three orders of magnitude energy efficiency gain at the same performance (iso-performance level) as CMOS.

Several superconductor logic families, such as Rapid single flux quantum (RSFQ), energy-efficient RSFQ (ERSFQ) \cite{b_[1]}, and adiabatic quantum-flux parametron (AQFP) \cite{b_[2]}, are introduced in the literature. However, the lack of dense memory, limited fanin/fanout, and deep pipeline nature of these logics requires support from advanced circuit methodologies. Much work on SCE library cell design and characterization, circuit simulation and margin, power and timing analysis, logic synthesis, and physical design, as well as biasing networks and dual clocking architectures, was performed as part of the IARPA C3 and SuperTools programs \cite{b_[4], b_[5]}. To show the feasibility of this technology, fully functional SCE-based microprocessor prototype chips have been demonstrated in RSFQ \cite{b_[6]} and AQFP logic \cite{b_[7]}.

The deeply pipelined nature of SCE logic makes it more suitable for applications with streaming data processes such as fast Fourier transform (FFT) or number theory transform (NTT). Therefore, the most process-intensive part of FHE, NTT, can benefit from SCE logic. Fully homomorphic encryption (FHE) is a cryptosystem that allows operations to be evaluated on encrypted data without decryption \cite{b_[8]}. It has been one of the most promising solutions that makes it possible to outsource computation and securely aggregate sensitive information from individuals, corporations, and government agencies.

With the potential development of large-scale quantum computers, existing public-key cryptography protocols (such as RSA and ECC) become insecure due to the quantum Fourier transform and Shor's algorithm \cite{b_Shore}. As a prime candidate for quantum-resistant encryption algorithms, FHE enables users and developers to perform computations directly on the ciphertexts, obtaining results identical to exact computations on plaintexts. After Gentry's first construction of fully homomorphic encryption \cite{b_[8]}, several research studies \cite{b_[9], b_[10], b_[11], b_[12], b_[13], b_[14]} have improved the efficiency of FHE schemes.

FHE schemes can be classified into different types based on the data types they support. The most common FHE schemes are the following.
\begin{enumerate}
    \item The Fast Homomorphic Encryption from Learning with Errors (FHEW) scheme \cite{b_FHEW, b_TFHE} works primarily on Boolean logic.
    \item The Brakerski-Gentry-Vaikuntanathan (BGV) and Brakerski-Fan-Vercauteren (BFV) schemes \cite{b_BGV, b_BFV} support modular arithmetic over finite fields.
    \item  The Cheon-Kim-Kim-Song (CKKS) scheme \cite{b_CKKS} relies on approximate computation.
\end{enumerate}

All these schemes are developed based on the mathematical hardness of ring learning with error, where noise is added during encryption and computations on the ciphertext. Therefore, FHE schemes require the ciphertext modulus $q$ to be large enough to accommodate noise growth without corrupting useful information.

There are two main categories of memory access algorithms that are used to execute the NTT operation efficiently: in-place and out-of-place memory access algorithms. The in-place memory access algorithm writes all intermediate results at the location from which they are read. Consequently, this approach necessitates a data rerouting network and random access memory to correctly read the two coefficients as a pair, accounting for accessing different coefficients across NTT stages. On the other hand, out-of-place memory access emerges as a good candidate, allowing intermediate results to be stored in different locations compared to their read locations \cite{b_zeming_Hicop}.

As a result, simplifying the data rerouting network and random access memory becomes feasible, offering potential benefits to SFQ circuits. Additionally, multiple memories are used to mitigate write-after-read (WAR) hazards resulting from distinct read and write locations. Regarding hardware implementation, we need butterfly units (BUs) to perform butterfly operations in different stages, the twiddle factor memory to store the required twiddle factors, and the coefficient memory to store all intermediate results.

This paper presents an FHE accelerator that uses superconductor SFQ logic and memory. Meanwhile, the shift-register-based memory replaces the random access memory, and the corresponding control logic is significantly simplified. More specifically, we will focus on designing and demonstrating an accelerator for performing the number-theoretic transform (NTT) operation, which is the critical computational task in nearly all FHE operations, including ring multiplication, key switch, mod switch, and bootstrap.

\section{Homomorphic Operation}
Most ongoing research efforts focus on developing optimizations for the CKKS scheme \cite{b_SuYang_RNS_CKKS, b_Ahmet_CKKS, b_coxHE_CKKS}. The operations in the FHE schemes are performed in the ring introduced in \cite{b_RLWE, b_RLWE_simplified}. The ring can be denoted as $R_q = \mathbb{Z}_q[x] / \phi(x)$, where $Z_{q}[x]$ contains polynomials with coefficients in the range of $[0, q)$ and $\phi(x)$ is a reduction polynomial, typically equal to $x^N+1$. 

The polynomial degree of $\phi(x)$ limits the degree of polynomials in $R_q$. For example, when $\phi(x)$ equals $x^4+1$, the maximum polynomial degree for a polynomial inside $R_q$ is $3$. In addition, ciphertexts are stored in the polynomial, where the degree of a polynomial is usually denoted as \textit{N} and ranges from \textit{1,024} to \textit{65,536}. The primitive computations over ciphertext are mainly composed of polynomial multiplication over a ring $R_q = \mathbb{Z}_q[x] /(x^{N}+1)$, where $q$ is the modulus of the coefficients of the polynomials and the maximum degree of polynomials in this ring is equal to $N-1$. Consequently, ciphertext addition and multiplication operations correspond to polynomial additions and multiplications.

Instead of directly computing polynomial multiplication by matrix multiplication, which results in a time complexity of $O(N^2)$, the NTT reduces the time complexity to $O(N\log(N))$. This reduction is achieved by evaluating the polynomial:
\begin{equation}
\label{EQ: original NTT problem}
    A(x) = \sum_{i=0}^{N-1} a_i x^i\text{ mod }q
\end{equation}
at different powers of $N$-th primitive root $\omega_N$ (that is, twiddle factor). The $N$-th primitive root is a value that satisfies $\omega_N^{i} \not= 1 \text{ mod }q$ for $i\in (0, N)$ and $\omega_N^N = 1 \text{ mod }q$. 
Based on the divide-and-conquer method, this original problem can be separated into two sub-problems, as expressed as:
\begin{equation}
    \begin{aligned}
        A_{r} = A(\omega_N^{r}) &= \sum_{i=0}^{N/2-1} a_{2i} \omega_{\frac{N}{2}}^{ri} + \omega_N^r \sum_{i=0}^{N/2-1} a_{2i+1} \omega_{\frac{N}{2}}^{ri} \\
        A_{r+\frac{N}{2}} = A(\omega_N^{r+N/2}) &= \sum_{i=0}^{N/2-1} a_{2i} \omega_{\frac{N}{2}}^{ri} - \omega_N^r \sum_{i=0}^{N/2-1} a_{2i+1} \omega_{\frac{N}{2}}^{ri} 
    \end{aligned}    
\end{equation}
where $\omega_N^{r+N/2} = -\omega_N^{r}$ based on the halving lemma of the twiddle factor. 
Additionally, the sub-problems can be further separated.
Next, during the merge step, we can apply radix-$2$ Cooley-Tukey (CT) butterfly operations to compute the values of $A_r$ and $A_{r+N/2}$, which consist of modular multiplication, modular addition, and modular subtraction. Therefore, the NTT operation requires $s = \log_2(N)$ stages, each containing $N/2$ number of CT butterfly operations. 

Following the divide-and-conquer method, in the first NTT stage, the CT butterfly operation can be expressed as:
\begin{equation}
    \begin{aligned}
        a_i &= a_i + \omega_2^0 a_{i+\frac{N}{2}} \\
        a_{i+\frac{N}{2}} &= a_i - \omega_2^0 a_{i+\frac{N}{2}}
    \end{aligned}
\end{equation}
where $i \in [0, N/2)$. 
Two coefficients are processed as a pair of coefficients: $(a_i, a_{i+N/2})$. Notice that the difference between coefficient indices inside a pair will decrease from $N/2$ in the first NTT stage to $1$ in the last NTT stage. In other words, when the index of a coefficient is in the binary representation $(i_{s-1} i_{s-2} \cdots i_1 i_0)$, the bit that is different between coefficient indices inside a pair will vary from $i_{s-1}$ in the first NTT stage to $i_0$ in the last NTT stage.

\section{FHE Accelerator}
\begin{figure}[!t]
    \centering
    \includegraphics[width=2.6in]{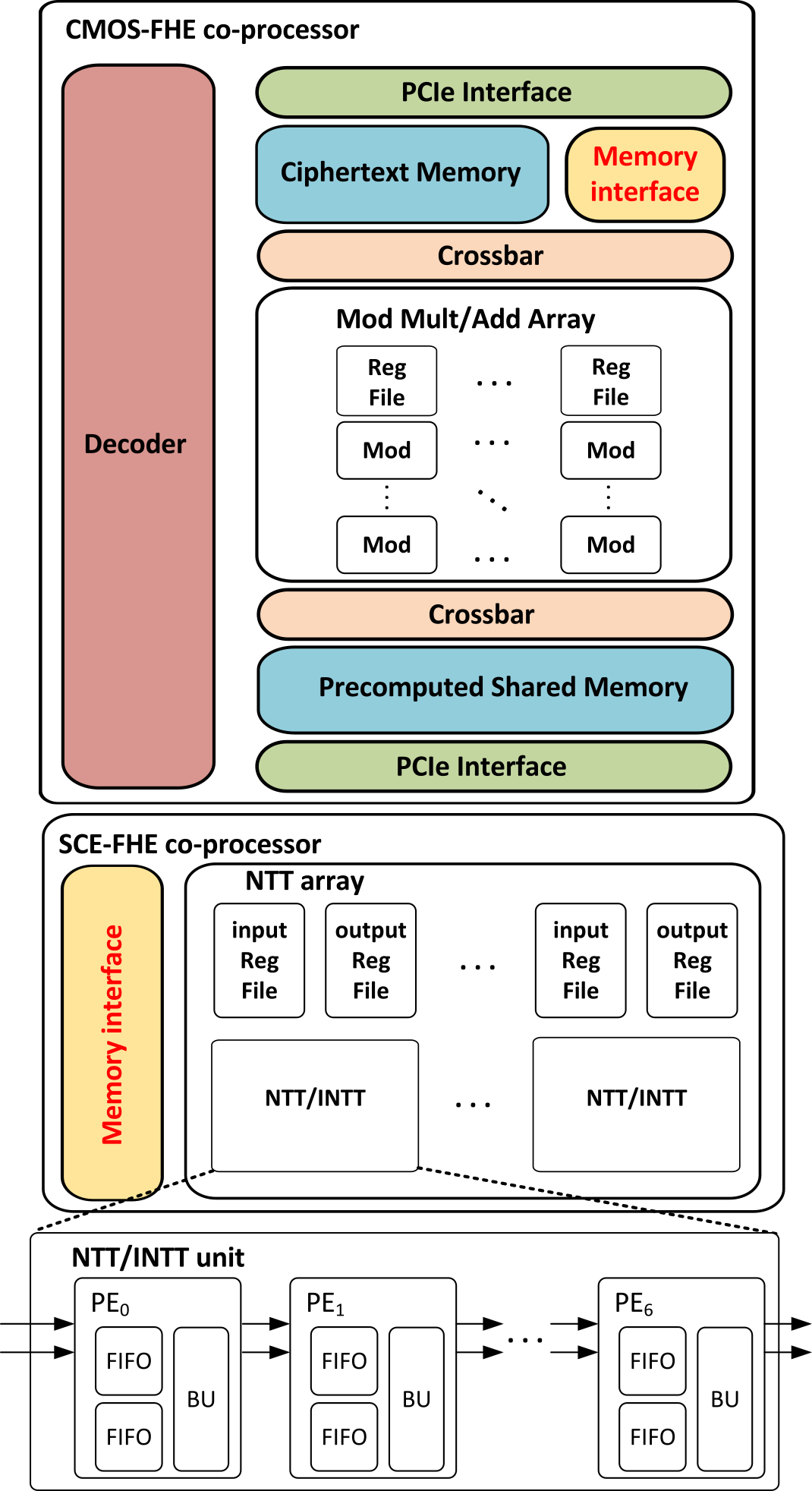}
    \caption{The High-level Diagram of the proposed FHE accelerator is demonstrated. CMOS and SCE are connected via the "Memory Interface" unit.}
    \label{FIG: The High-level Diagram of FHE accelerator}
\end{figure}

To develop an FHE accelerator using superconductive SFQ logic and memory, we propose the architecture depicted in Fig.~\ref{FIG: The High-level Diagram of FHE accelerator}. The design comprises two coprocessors: a CMOS-FHE coprocessor operating at room temperature and an SCE-NTT coprocessor operating at cryogenic temperatures. The CMOS-FHE coprocessor facilitates data transfer between the host and the SCE-NTT coprocessor since dense on-chip SFQ memory with random access has not yet been realized. The CMOS-FHE coprocessor can also perform modular additions and multiplications to accelerate various auxiliary operations in an FHE system. The SCE-NTT coprocessor is employed to speed up the NTT and inverse NTT (iNTT) operations utilizing the ultra-high-frequency switching nature of the SFQ logic. 

\begin{table}[!t]
\caption{Operation Decomposition of CKKS Scheme For Different coprocessors \label{TABLE: Operation Decomposition of CKKS Scheme}}
\centering
\scriptsize
\begin{tabular}{|p{1.3cm}|p{1.2cm}|p{1.6cm}|p{1.2cm}|}
\hline
Operations & Host & CMOS coprocessor & SCE-NTT coprocessor \\ \hline
Mod Up/Down & Data Transfer & Decode, Modular Mult, Modular Add/Sub & NTT, INTT \\ \hline
Relinearization & Data Transfer & Decode, Modular Mult, Modular Add/Sub& NTT, INTT \\ \hline
Rescale & Data Transfer & Decode, Modular Mult, Modular Add/Sub & None \\ \hline
Homomorphic Add/Sub & Data Transfer & Decode, Modular Add/Sub & None \\ \hline
Homomorphic Mult & Data Transfer & Decode, Modular Mult, Modular Add/Sub & NTT, INTT \\ \hline
\end{tabular}
\end{table}

In the CKKS scheme, operations can be broken down into modular addition/subtraction, modular multiplication, NTT, and INTT operations. Table \ref{TABLE: Operation Decomposition of CKKS Scheme} delineates the specific operations handled by each coprocessor. Specifically, in this seedling project, our focus is on designing and demonstrating an accelerator (NTT unit) for performing the NTT and INTT operations, which are critical computational tasks in various FHE operations, including Homomorphic Multiplication, Mod Up/Down, and Relinearization.

\section{Efficient NTT Implementation}
Figure \ref{FIG: The NTT-128 architecture design} provides an overview of our NTT design with 128 points (NTT-128) using SFQ circuits. The design comprises seven processing elements (PEs), with each PE handling butterfly operations for one NTT stage. The data flow in the proposed NTT-128 design follows a pipeline structure, progressing from the first PE to the last PE. Each PE comprises two coefficient memories, each with \textit{N} slots operating in a ping-pong manner to store all intermediate results to avoid WAR hazards since the read and write addresses of the coefficients are different. Additionally, a BU is included to compute butterfly operations within an NTT stage. Furthermore, two twiddle-factor queues are incorporated to store precomputed values. 

\subsection{NTT Architecture}
\begin{figure}[!t]
    \centering
    \includegraphics[width=3.4in]{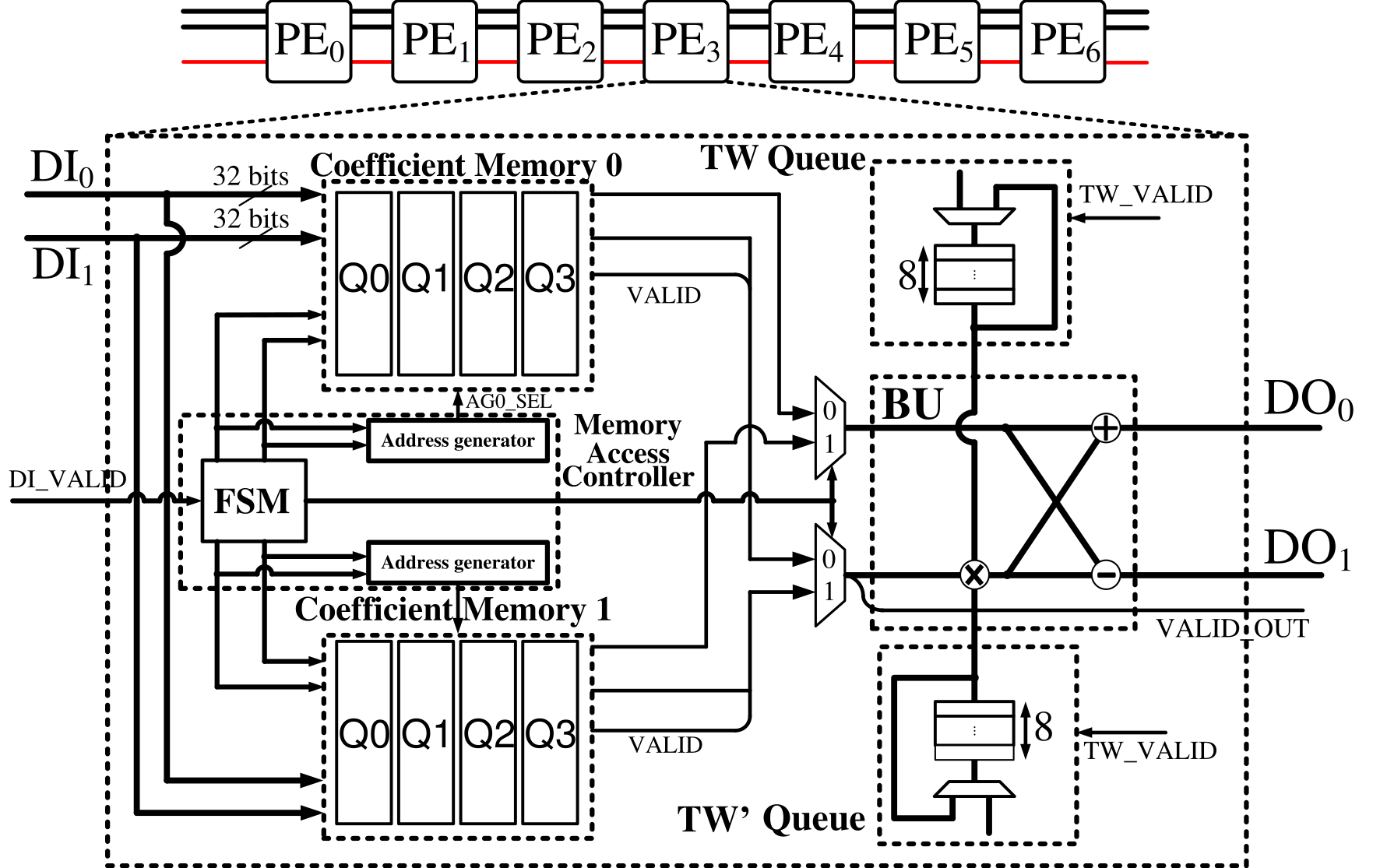}
    \caption{The NTT-128 architecture design is demonstrated. The design has 7 PE stages, each of which comprises 32-bit BU and memory units.}
	\label{FIG: The NTT-128 architecture design}
\end{figure}

The proposed NTT operates in a fully pipelined manner, while the two coefficient memories operate in a ping-pong manner. During computation, one coefficient memory is used to store intermediate results from the primary input or previous PE, and the other coefficient memory provides data from previous NTT operations to the BU for butterfly operations. Meanwhile, the twiddle factor (TW) queues provide the corresponding TWs to the BUs, and the pipelined BU executes butterfly operations for an NTT stage, generating the results as the primary output of a PE.  

\begin{figure}[!t]
    \centering
    \includegraphics[width=3.3in]{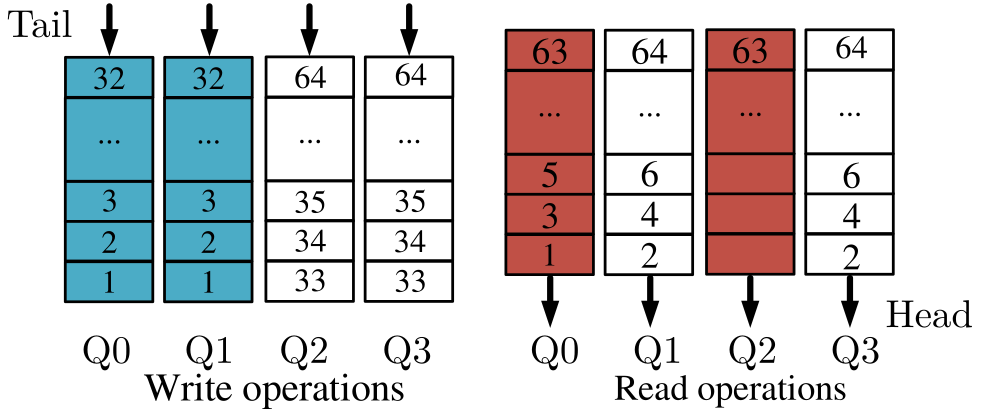}
	\caption{The memory access pattern of coefficient memory}
	\label{FIG: The memory access pattern of coefficient memory}
\end{figure}

Our design utilizes the first-in-first-out queues (FIFO), which are implementable using shift registers, to instantiate coefficient and twiddle factor memories. This approach, with the proposed memory access pattern, eliminates the need for complex random access memory (RAM) in the SFQ circuit. Figure \ref{FIG: The memory access pattern of coefficient memory} illustrates the read and write sequences of the coefficient memory. This pattern will be applied consistently across all coefficient memories in various PEs.

Based on the property of the FIFO, the write operation can only be performed at the tail of the queues, whereas the read operation can only be performed at the head of the queues. Each coefficient memory consists of four queues that can store or generate two coefficients in one clock cycle. The numbers in Figure \ref{FIG: The memory access pattern of coefficient memory} denote the sequence for read and write operations. For example, we will write two coefficients to queues 0 and 1 in the first clock cycle. 
Next, the corresponding coefficients will be shifted to the heads of the queues.

To establish the connection between coefficient indices and memory locations, we use the binary representation of the coefficient index \textit{i} to denote the memory location of the coefficient $a_i$.
Assume that two coefficients are provided to the first PE in the natural order (e.g., $a_0$ and $a_1$ in the first cycle, followed by $a_2$ and $a_3$). Following the write access pattern shown in Figure \ref{FIG: The memory access pattern of coefficient memory}, the first half of the coefficients will be stored in queues 0 and 1, and the last half will be stored in queues 2 and 3. We can, therefore, derive that the coefficient $a_i$ is stored at the memory location:
\begin{equation}
    (\bm{i_6} i_5 i_4 i_3 i_2 i_1 \bm{i_0})_b
    \label{EQ: memory address in stage 1}
\end{equation}
where the most significant bit $(i_0)$ and the least significant bit $(i_6)$ are in bold since they serve as bank enables for four queues. The rest of the bits ($i_5$ to $i_1$) are the corresponding addresses of a coefficient inside a queue. As these storing operations are performed, $a_0$ and $a_1$ in queues 0 and 1 are shifted to the heads of the queues. Therefore, the coefficient $a_i$ is stored in the location specified in equation \eqref{EQ: memory address in stage 1}. 

After all coefficients are stored inside these four queues, we start the read operations following the sequence shown in Figure \ref{FIG: The memory access pattern of coefficient memory}. During read operations, two coefficients stored in queues 0 and 2 or 1 and 3 are obtained in one clock cycle. 
Therefore, the indices of coefficients within a coefficient pair differ in $i_6$, meeting the requirement for inputs to butterfly operations in the first NTT stage. The corresponding bits $(i_5i_4i_3 i_2 i_1i_0)_b$ of the coefficient indices within a pair of coefficients will automatically be incremented according to the property of the FIFO. While coefficient memory '0' generates coefficients for inputs of the BU, coefficient memory '1' stores coefficients for the next NTT operation.

The results of butterfly operations in the first PE will be applied to the second PE. The input sequence of the coefficient for the second PE is identical to the read sequence in the first PE since the BU is a pipelined design. Consequently, after applying the same write sequence as shown in Figure \ref{FIG: The memory access pattern of coefficient memory} for coefficient memory in the second PE, the coefficient $a_i$ is stored at the memory location:
\begin{equation}
    (\bm{i_5} i_4 i_3 i_2 i_1 i_0 \bm{i_6})_b
\end{equation}
where coefficient pairs are stored in queues 0 and 1, followed by queues 2 and 3. Therefore, $i_6$, which is the different bit inside coefficient indices inside a pair in the first stage, is moved to the least significant bit of the addresses. Similarly, following the same read sequence, two coefficients with different values on $i_5$ will be obtained with the read address $(i_4 i_3 i_2 i_1 i_0 i_6)_b$. 
Based on the retrieved coefficient pairs, we can perform butterfly operations in the second NTT stage. In addition, we can derive the memory layout in the third PE as,
\begin{equation}
    (\bm{i_4} i_3 i_2 i_1 i_0 i_6 \bm{i_5})_b
\end{equation}
This process can be repeated until the last PE. The efficient computation of a 128-point NTT operation is achieved through the proposed memory access pattern utilizing seven PEs, with each PE performing butterfly operations for a specific NTT stage.

Since the read-access pattern for the coefficients is fixed across different PEs, the sequence of TWs will also be fixed and easily implemented using a wrap-around FIFO, where the input and output are connected.
Additionally, based on the property of the divide-and-conquer method, the number of required TWs will vary between different PEs, with the last PE requiring the largest number of TWs (that is, 64). Furthermore, another TW queue is denoted as TW' (also called TWP) in Figure \ref{FIG: The NTT-128 architecture design}. The TW' queue stores precomputed values corresponding to twiddle factors within the TW queue, resulting in the same number of memory slots as the TW queue. These precomputed values are used to save hardware area for the modular multiplication unit.

\subsection{Butterfly Unit Design}
The BU comprises a modular multiplier, a modular subtractor, and a modular adder. For modular multiplication $(C = A \times B \text{ mod }q)$, we have three prevailing types of design: Shoup modular multiplier, Montgomery modular multiplier, and Barrett modular multiplier. The details of each design can be found in \cite{b_Shoup, b_zeming_iterative_BMM, b_zeming_iterative_MMM}. As the values of \textit{B} in the butterfly operations are precomputed twiddle factors, we can take advantage of the Shoup modular multiplier.   
This approach requires a smaller hardware area, aided by the value of \textit{B'}, which is associated with both \textit{B} and \textit{q}. These \textit{B'} values are stored in the TW' queue.

Since the RSFQ circuit is deep-pipelined, meaning that every logic cell requires a clock signal, we need to insert path-balancing DFFs to balance the data path for each cell. Based on our result from \cite{b_[5]}, as the circuit size grows, the number of path-balancing DFFs can take more than 60$\%$ of the whole design. Furthermore, since each cell requires a clock signal and the RSFQ circuit requires a specific cell called a splitter to propagate the corresponding signal to multiple fanouts, we need a large clock tree to distribute this signal. The large amount of path-balancing DFFs makes the situation even more complicated \cite{razmkhah2024challenges}. Therefore, our preferred design is an architecture that performs an operation in parallel and has a smaller circuit depth.

\begin{figure}[!t]
    \centering
    \includegraphics[width=3.5in]{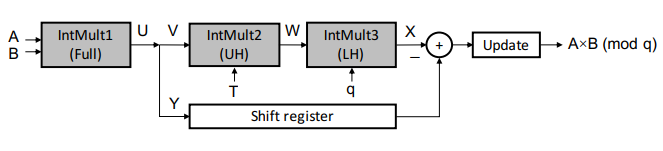}
	\caption{The schematic of Barrett modular multiplier}
	\label{FIG:Barrett}
\end{figure}

\begin{figure}[!t]
    \centering
    \includegraphics[width=3.25in]{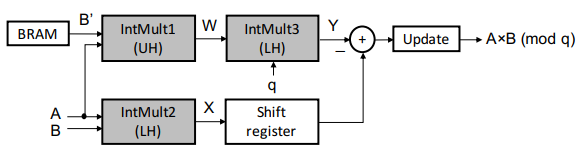}
	\caption{The schematic of Shoup Modular Multiplier}
	\label{FIG:Shoup}
\end{figure}

The Montgomery modular multiplier requires Montgomery conversion before it can perform a modular operation, resulting in a specialized converter in the BU. Hence, we select between the Barrett modular multiplier and Shoup modular multiplier, whose schematics are shown in Fig.~\ref {FIG:Barrett} and Fig.~\ref {FIG:Shoup}.

Based on the schematic, the Barrett modular multiplier comprises a multiplier, two half multipliers, path-balancing DFFs with a depth of two half multipliers, and a modular subtractor. The Shoup modular multiplier comprises three half multipliers, path-balancing DFFs with a depth of one-half multiplier, and one modular subtractor. The circuit depth is two half multipliers and one modular subtractor. The synthesis result for the Barrett and Shoup modular multiplier is shown in Table~\ref{tab:modular_multiplier}.

\begin{table}[!t]
\caption{Comparison between Barrett and Shoup Mod multiplier\label{tab:modular_multiplier}}
\centering
\begin{tabular}{|p{1.4cm}|p{1.2cm}|p{1.1cm}|p{1.1cm}|p{1.2cm}|}
\hline
\textbf{Design}& \textbf{Logic cell Count} & \textbf{Splitter Count} & \textbf{DFF Count} & \textbf{JJ Count}  \\ \hline
Barrett Mod Multiplier & 31,848 & 108,738 & 48,060 & 1,342,704 \\ \hline
Shoup Mod Multiplier   & 18,583 & 56,324 & 18,433 & 664,873   \\ \hline
\end{tabular}
\end{table}

As shown in Table~\ref{tab:modular_multiplier}, the JJ count for the Shoup is around half of the Barrett modular multiplier due to the small number of path-balancing DFFs and Splitters. Therefore, we use the Shoup modular multiplier in our design.

The Kogge-Stone adder fits our design best because it only requires log(N) levels to finish the design, which saves a large area for our design. In comparison, the subtractor is just one more layer to convert the number to its two's complement.

\begin{figure}[!t]
    \centering
    \includegraphics[width=3.5in]{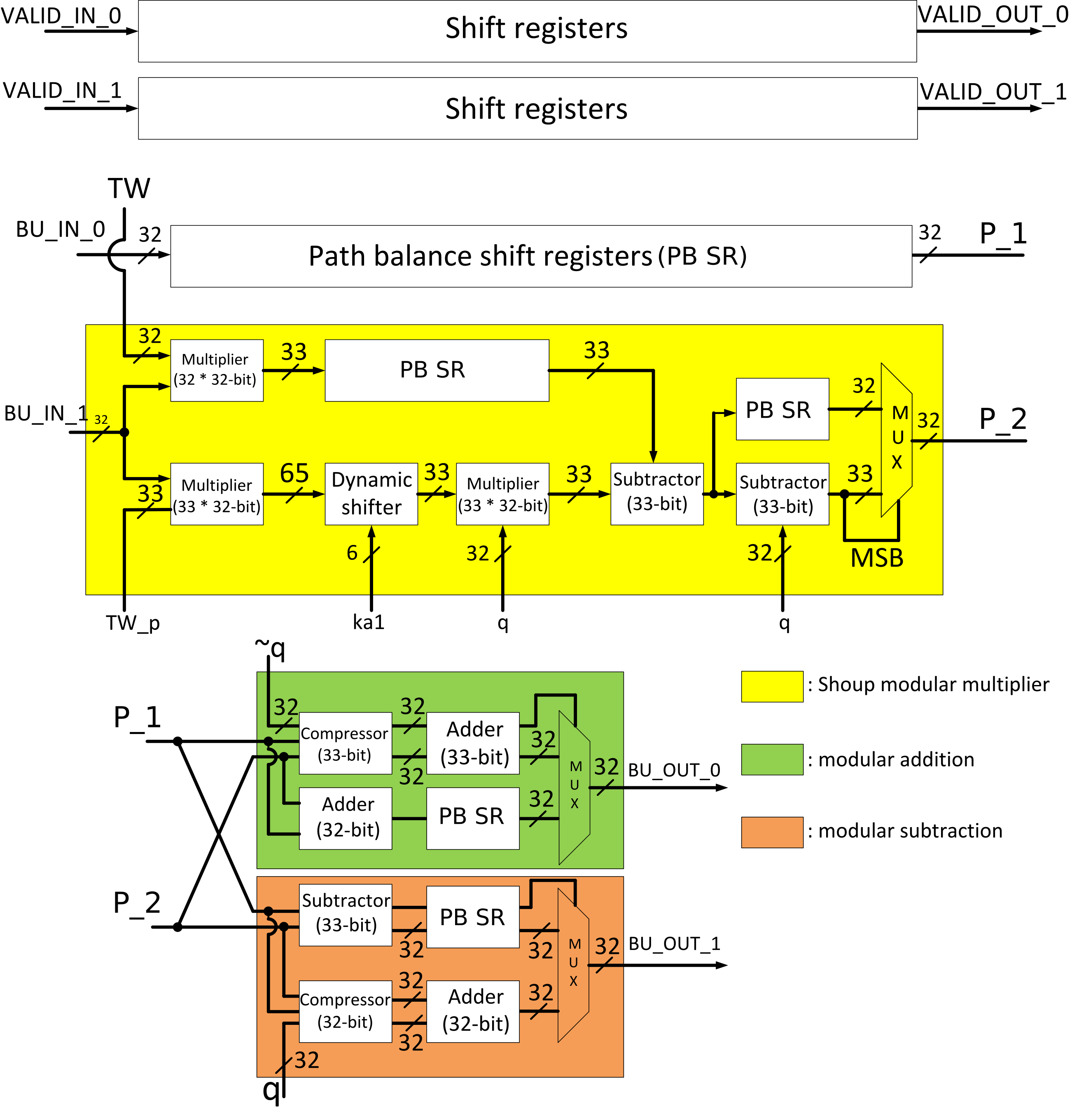}
	\caption{Schematic of the Butterfly unit with its sub-components. }
	\label{FIG:Butterfly unit}
\end{figure}

As shown in Fig.~\ref{FIG:Butterfly unit}, the yellow block is the Shoup modular multiplier, the green block is the modular adder, and the orange block is the modular subtractor. The synthesis results for BU and the memory block are shown in the Table.~\ref{tab:whole result}.

\begin{table}[!t]
\caption{Post synthesis result for whole NTT design. The clock cycle for the design is 29.4 ps, and the throughput is 34 GHz, which translates to 531M (mega) NTT/sec.~\label{tab:whole result}}
\centering
\begin{tabular}{|p{1.5cm}|p{1.2cm}|p{1.2cm}|p{1.3cm}|}
\hline
\footnotesize
\textbf{Component} & \textbf{Logic Cell Count} & \textbf{Total JJ } & \textbf{Latency}\\ \hline
BU for each PE & 19,838 &969,774 & 79 cycles \\ \hline
Total Datapath & 138,866 &6,788,446 & 79 cycles\\ \hline
Memory block for the last PE (max) & - & 182,474 & 69 cycles\\ \hline
Total memory & - &1,099,309 & 69 cycles\\ \hline
Total design & - &7,887,755  & 1,036 cycles\\ \hline
\end{tabular}
\end{table}

\section{RSFQ Circuit Implementation}
Such a large circuit synthesis needs a complete cell library with compound functions such as multiplexors and a one-bit full adder. A new cell library with the MIT LL SFQ5ee process \cite{MITLL_tolpygo2015} in mind was designed using a hierarchical and parameterized approach. All component values are designated as parameters. These values are scaled according to the parameters of the fabrication process. This approach allows frequently utilized circuit structures, or "components," to be defined as distinct subcircuits, promotes reusability and scalability, facilitates library retargeting to other fabrication processes, accelerates the cell design process, and optimization to subcomponents automatically propagates through the entire library.

\begin{figure}[!t]
	\centering
	\includegraphics[width=3.3in]{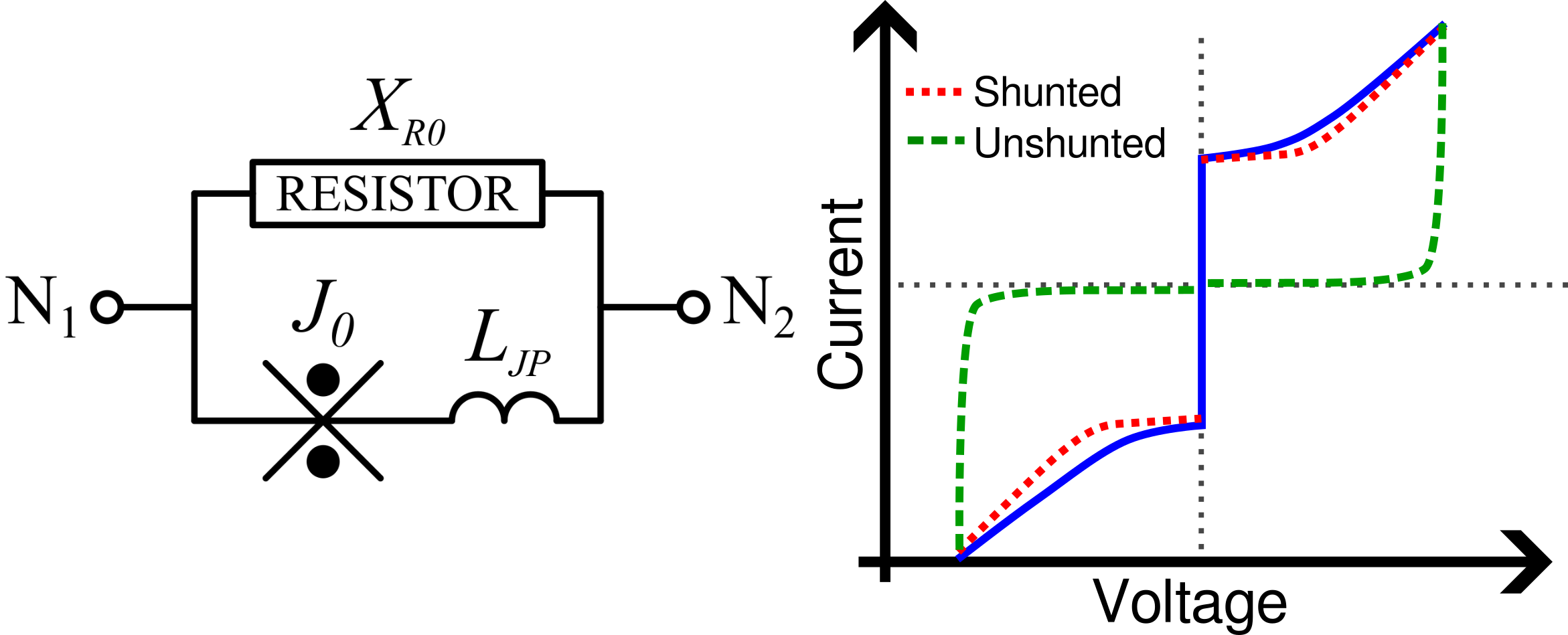}
	\caption{Parameterized shunted Josephson junction and its I-V characteristics. The shunt value affects the hysteresis in the I-V curve.}
	\label{fig:su_comp_jjs}
\end{figure}

\paragraph{Cell Library Structure}
The RSFQ cells are constructed using smaller parameterized components known as subcircuits. Each subcircuit comprises basic circuit elements such as inductances and resistances. The switching element in the RSFQ circuits is JJ. The schematic of a JJ with parasitic parameters is shown in Fig.~\ref{fig:su_comp_jjs}. The parameters that are controlled for cell design are demonstrated in Table~\ref{tab:su_comp_jjs}. The program gets the parameters and, based on that, forms the cell library that comprises global parameters coming from the fabrication process, components that house all the smaller subcircuit components routinely used in cells, base cells that store all the logic, interface, and buffer cells, standard cells that contain cells with standard connections, PTL cells containing cells with Passive Transmission Line (PTL) interfaces designed to connect to PTLs and includes integrated PTL transmitter and receiver circuits.

\begin{table}[ht]
\centering
\caption{Junction parameters.}
\label{tab:su_comp_jjs}
\footnotesize

\begin{tabular}{ |p{0.15\linewidth}|p{0.25\linewidth}|p{0.40\linewidth}| }
\hline
\textbf{Parameter} & \textbf{Default Value} & \textbf{Description} \\ 
\hline
\texttt{Ic}		& \texttt{GLOBAL\_$I_{C0}$}& Junction critical current [A]. \\
\hline
\texttt{BetaC}	& \texttt{GLOBAL\_$\beta_C$} & Junction Stewart-McCumber parameter. \\
\hline
\texttt{JJarea}	& \texttt{Ic/FABPROC\_Jc} & Junction area [$m^2$]. \\
\hline
\texttt{R0}		& \texttt{$\sqrt{\frac{\beta_C*\Phi_0}{2*\pi*I_C*C_s*A}}$} & Junction shunt resistor [$\Omega$]. \\
\hline
\texttt{Ljp}		& \texttt{GLOBAL\_LJP} & Junction parasitic inductor [H]. \\
\hline
\hline
\end{tabular}
\end{table}

\begin{figure}[!t]
	\centering
	\includegraphics[width=2.5in]{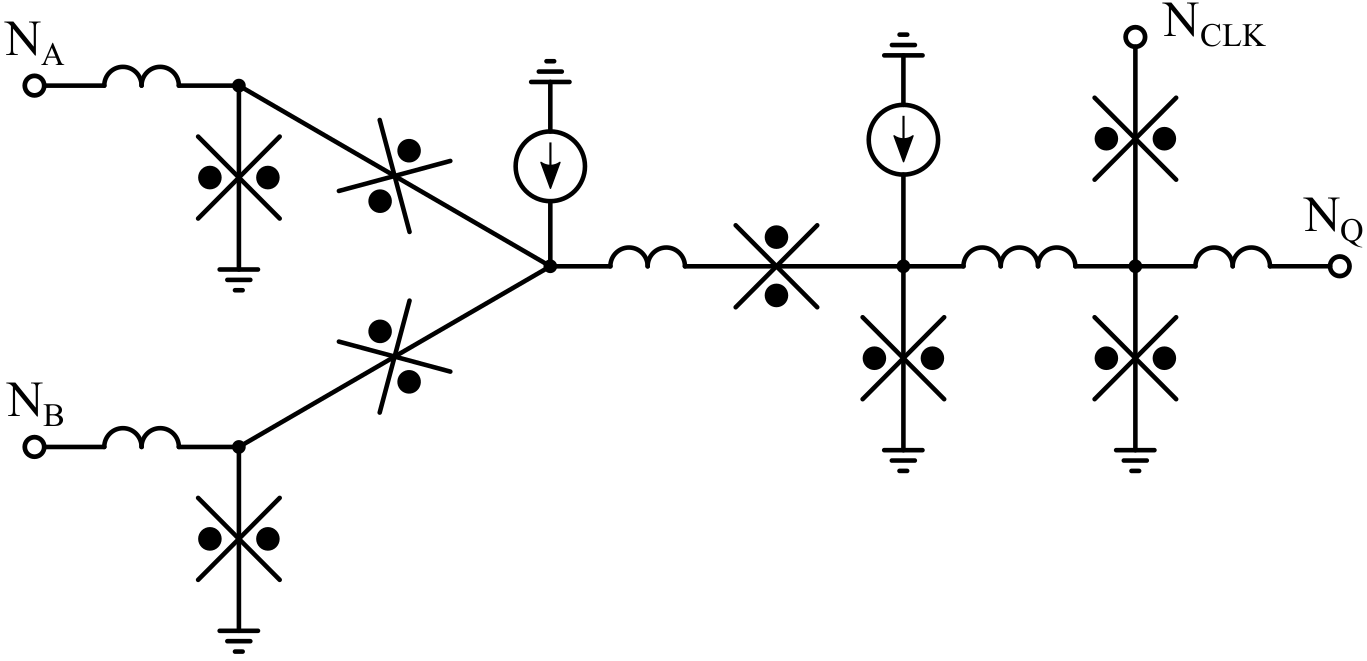}
	\caption{OR2\_base cell Original schematic.}
	\label{fig:su_or2_base}
\end{figure}

\begin{figure}[!t]
	\centering
	\includegraphics[width=2.6in]{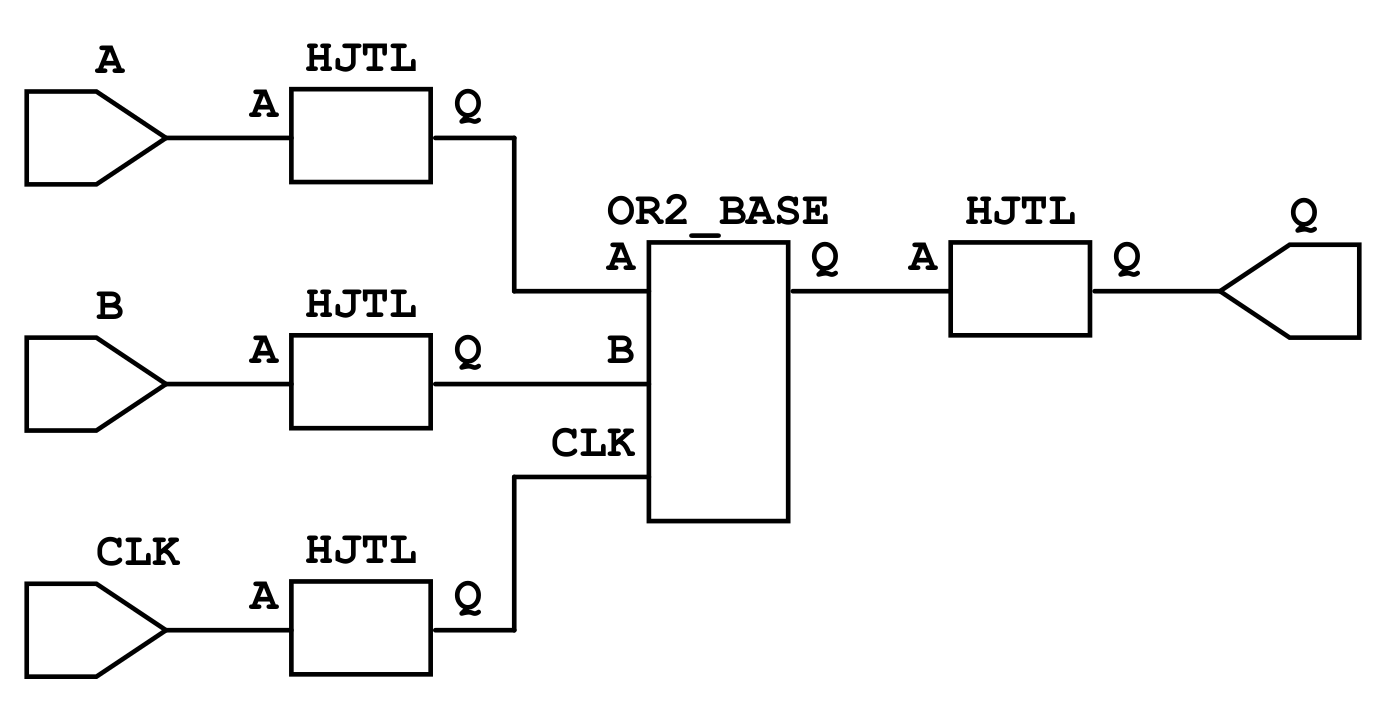}
	\caption{A base cell \texttt{OR2\_BASE} with standard interconnects. For PTL interconnects, the HJTL cells are changed with PTLRx at the input terminals and PTLTx at the output.}
	\label{fig:su_or2_standard}
\end{figure}

An OR base cell structure is shown in Fig.~\ref{fig:su_or2_base}. Standard cells are designed for direct connection to each other. The input and output pins incorporate integrated standard transmitter and receiver cells called the HJTL (half Josephson transmission line). All standard cells use identical transmitters and receivers, ensuring stable interconnects, as demonstrated in Fig.~\ref{fig:su_or2_standard}. Our cell library contains over 50 standard cells, with all base logic cells operating at 50 GHz.

\paragraph{PTL interconnects}
Passive transmission lines are implemented as striplines with ground planes above and below each conductor. To decrease the layout footprint of the interconnects, the characteristic impedance has been increased from the previous \SI{5.4}{\ohm} presented in \cite{b_[5]} to \SI{11}{\ohm} in the new library to increase the routing density. This results in the width of the signal line changing from a width of \SI{4.6}{\micro\meter} to \SI{2}{\micro\meter}, effectively doubling the routing capacity for a given area. 

\section{SRM and Controller Design}
As stated, memory implementation in SFQ technology is a challenge. Therefore, the architecture design aimed at using SRM is more feasible for SFQ design and meets the timing closure confirmed by behavioral and circuit simulations up to 34~GHz. Fig.~\ref{FIG: The NTT-128 architecture design} shows a high-level diagram of a single PE stage, which includes the BU, memory subblocks, a controller, and some glue logic. Notice that there is an SRM for storing coefficients, circular shift register memories (CSRM) for storing TW and TW's data, a memory access controller (MAC) to orchestrate the data movement, and glue logic to tie all the main memory components together, including asynchronous merger-based MUXes and delay elements.

The memory subblock is clocked at 34~GHz (29.4~ps clock period) and has a control signal to commence loading of TW data (TW\_LD) and notify that the last TW data has been loaded (TW\_RDY). There are five 32-bit input ports, two for coefficient input data (DI0, DI1), two for the Valid bit for data (VI0, VI1), and loading initialization of 32-bit data for TW CSRM (TWI). The input port for loading the initialization of 32-bit data for TW's CSRM needs 33 bits.

\begin{figure}[!t]
  \centering
  \includegraphics[width=3.5in]{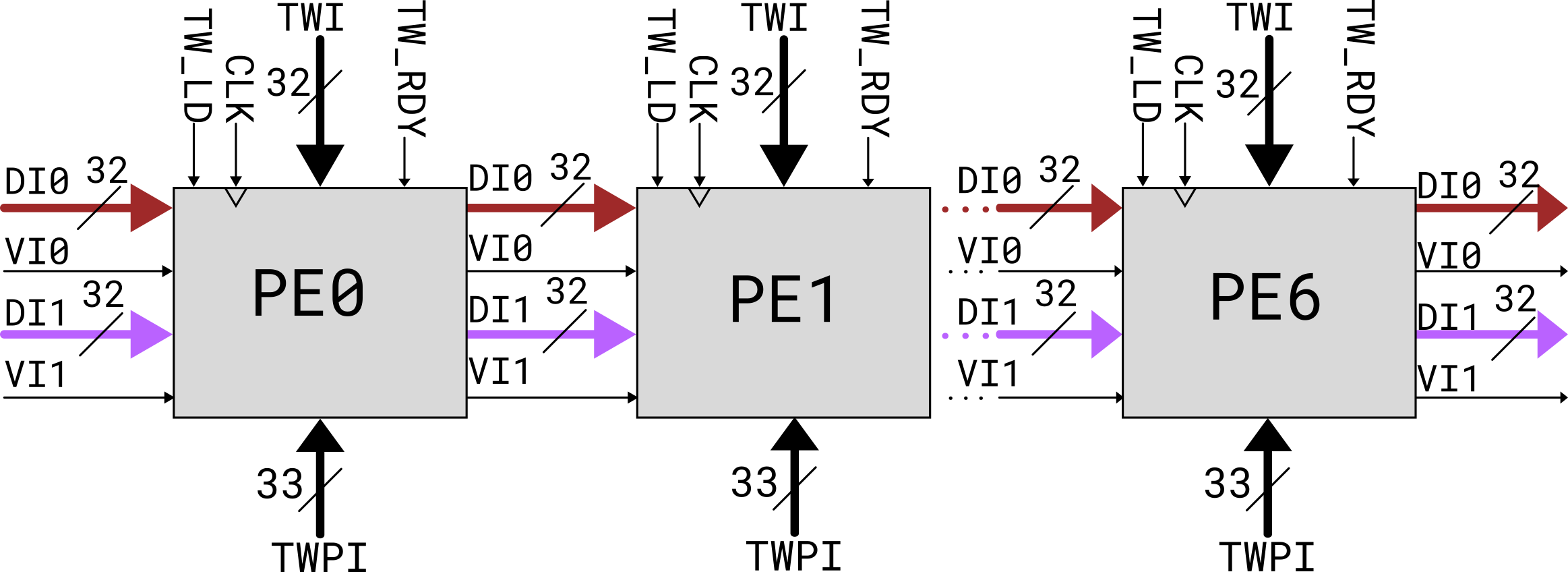}
  \caption{Top-level I/O of memory/processing element.}
  \label{fig:mem-io}
\end{figure}

The connections between the PEs are shown in Fig.~\ref{fig:mem-io} and the primary I/O control and loading signals. The difference between each memory subblock design is the TW memory size, starting with a 1-stage CSRM for PE0 to a 64-stage CSRM for PE6 ($\text{CSRM stage size}=2^i$ for $\text{PE}_i$). The latency of this memory (clk-to-q) is 7.5 cycles (219~ps), the output spread is $\pm$2.5~ps, and setup and hold times are zero and 5~ps, respectively. The coefficient memory has two banks, and the bank size is 4$\times$32$\times$32 bits. TW memory has two banks, and each bank's maximum length is 64 $\times$ 33~bits.

\subsection{Full-Custom Design Infrastructure for Memory}
\label{sec:memory-infrastructure}
\begin{figure}[!t]
  \centering
  \includegraphics[width=\linewidth]{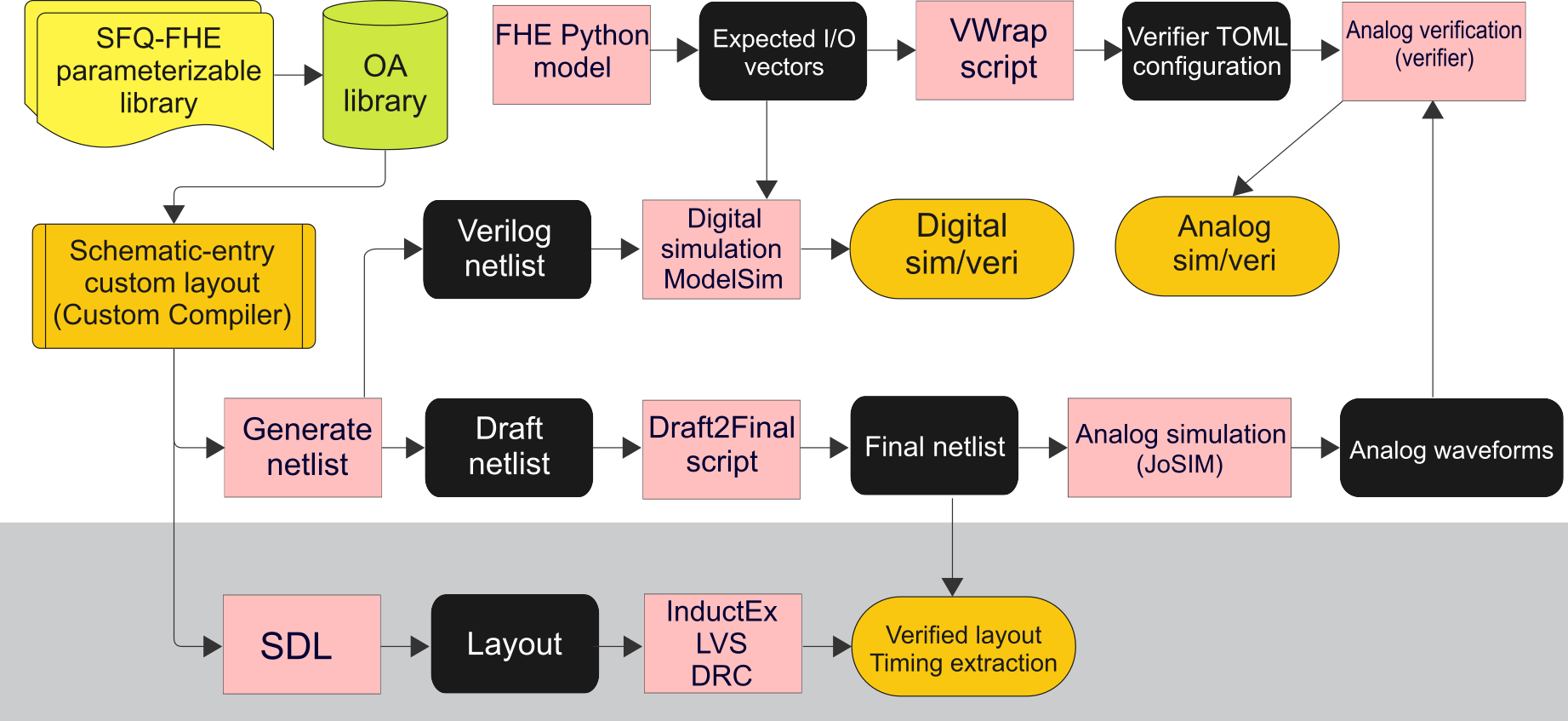}
  \caption{Memory full-custom design flow.}
  \label{fig:mem-full-custom-flow}
\end{figure}
The memory subblock of each PE is designed using a full custom design flow illustrated in Fig.~\ref{fig:mem-full-custom-flow}. We use Synopsys Custom Compiler for gate-level, full-custom design of the memory system, which includes the hierarchical management of the memory subblock and components. The design is primarily done at the gate level using a schematic entry tool with access to the cell library database. Netlists are generated from schematics and processed for analog and digital simulations. The functionality of the design was verified by comparing the analog simulation results from the JoSIM simulator to the high-level Python model.

\subsection{Full-Custom Design of SRM and MAC}
\label{sec:full-custom-mem}
The design of the memory subblock of the NTT processing element comprises four main subcomponents: coefficient SRM, twiddle factor CSRM, MAC, and Glue Logic. 

\subsubsection{Coefficient SRM}
\label{sec:srm}
\begin{figure}[!t]
  \centering
  \includegraphics[width=1.7in]{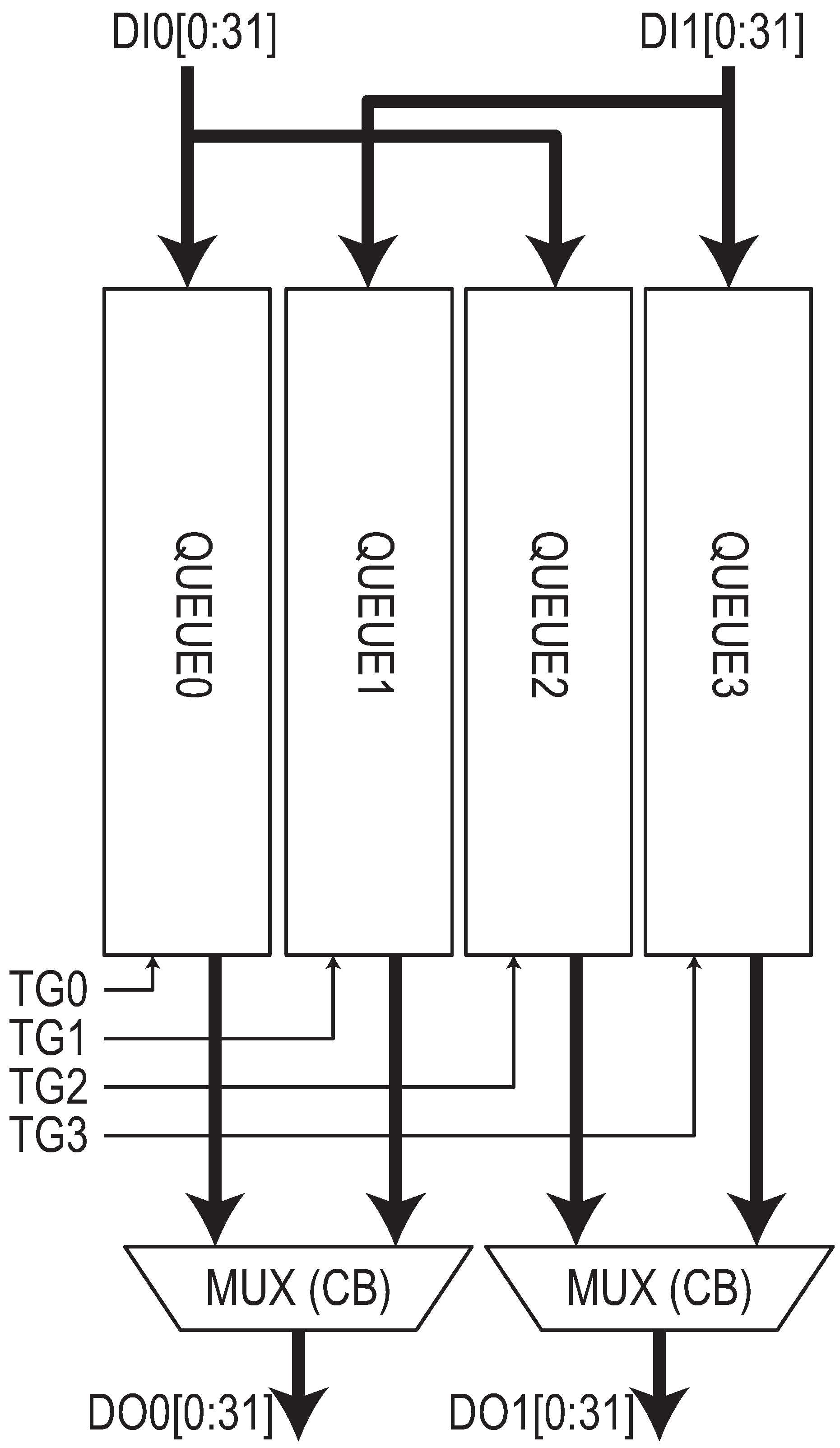}
  \caption{Memory coefficient bank.}
  \label{fig:mem-cm}
\end{figure}

As seen in Fig.~\ref{FIG: The NTT-128 architecture design}, two banks of coefficient SRMs are denoted as CM0 and CM1. When one bank is in the "read" mode, the other is in the "write" mode. When a bank is filled with data, it will switch to the "read" mode to start reading the data into the BU of the PE. Each bank comprises 4 SRM-based queues. Each queue is a 32-stage 32-bit SRM. A simplified diagram of the memory bank is shown in Fig.~\ref{fig:mem-cm}. Note that the term "queue" means that a queue is implemented as a standard shift register where data is loaded from the tail end, and on each shift cycle, data is shifted towards the head end of the shift register. Another important distinction is that the SRMs are "triggered" by the MAC at appropriate times rather than a system clock that always clocks the SRM on every cycle. So, what traditionally appears as a "clock" input for the SRM will, from now on, be referred to as the trigger (TRIG) input of the SRM.

\begin{figure}[!t]
  \centering
  \includegraphics[width=3.3in]{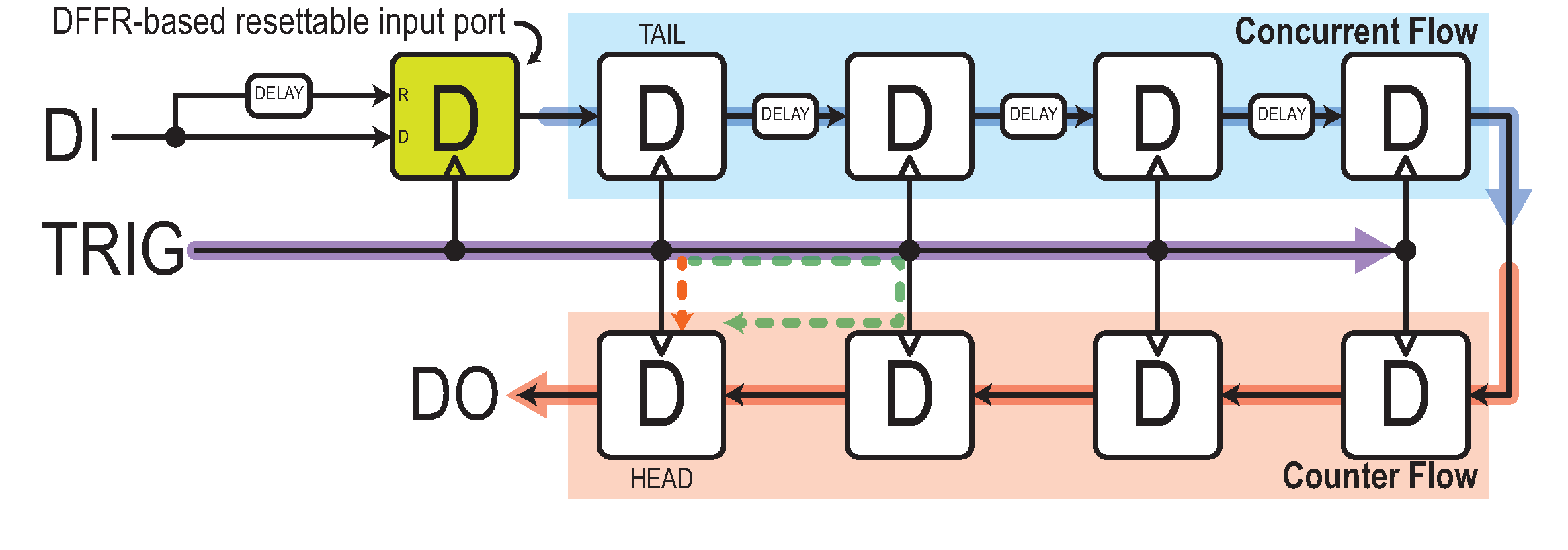}
  \caption{Coefficient queue, SRM implementation (bit-slice). The green dashed line corresponds to the critical performance limiting the path of a trigger pulse to shift data into the head DFF of the SRM. The orange dashed line denotes the shortest path of a subsequent clock pulse that reads out the head DFF. The absolute difference between green and orange paths determines the maximum clock frequency of the SRM.}
  \label{fig:mem-srm}
\end{figure}

The inputs to the coefficient SRM banks are broadcast to all possible destinations, whether they are the correct destination for that particular cycle or not. That is, \textbf{DIO} data is broadcast to QUEUE0 and QUEUE2, whereas \textbf{DI1} data is broadcast to QUEUE1 and QUEUE3. To DEMUX the broadcasted inputs appropriately, we designed a self-resettable write port comprising a resettable DFF (DFFR). This mechanism is illustrated in Fig.~\ref{fig:mem-srm} with the DFFR highlighted in green. The coefficient SRM consists of dual-flow clock-based shift registers with a write port based on a self-resettable DFFR. For each PE, there are two banks of coefficient memories. Each bank comprises four queues of 32-stage SRMs that are appropriately triggered during the read/write operations of the NTT, governed by the MAC.

Fig.~\ref{fig:mem-srm} shows a bit slice of the SRM. We extended the length of the bit slice to 32 stages (32 DFFs). The 32-stage bit-slice design is replicated 32 times to expand to 32 bits, and the incoming trigger signal goes through a FO4-FO4-FO2 fanout tree to provide the trigger to all 32 bits. The outputs are routed to merger-based asynchronous MUXes as illustrated in Fig.~\ref{FIG: The NTT-128 architecture design},\ref{fig:mem-cm}. Since only the appropriate queues are triggered for readout, data input collisions at the merger gates will not happen.

\subsubsection{Twiddle Factor CSRM}
\label{sec:csrm}
Twiddle Factor CSRM is a dual-flow clock-based shift register with feedback for circulating Twiddle Factor data. Each PE has two banks: one for standard Twiddle Factor data and one for Twiddle Factor prime data. There are two Twiddle Factor memories: the normal TW storing 32-bit data and the TW prime (TWP) storing 33-bit data. Both memories are CSRM, which keep the data circulating and enable initialization/loading. CSRMs use dual-flow clocking for easier feedback loop completion, and a non-destructive memory (NDRO) controls data loading to prevent data collision and spurious outputs. Data rotates through the CSRM until the NTT operation is completed, with the feedback loop potentially limiting performance. The CSRM size varies by PE, with the stage size for $\text{PE}_i$ being $2^i$.

\begin{figure}[!t]
  \centering
  \includegraphics[width=3.3in]{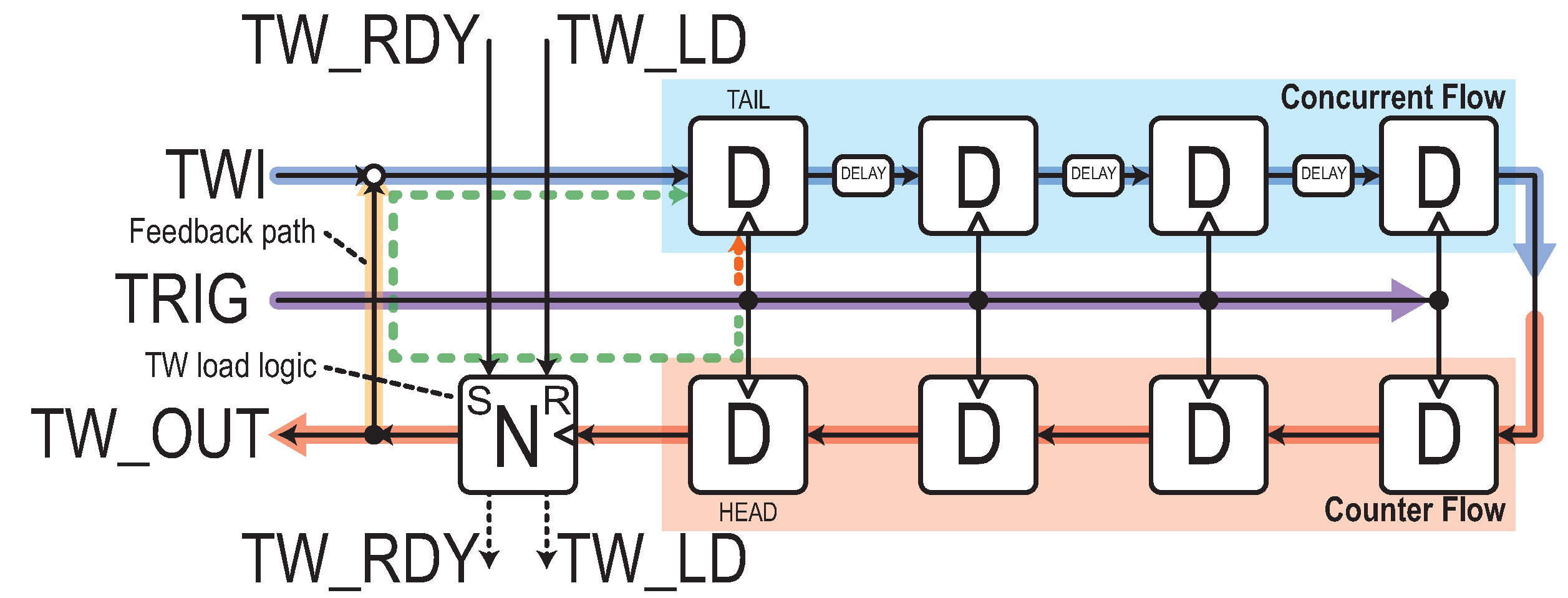}
  \caption{TW memory, CSRM implementation (bit-slice). The path it takes for a trigger clock to read out the head DFF and for that corresponding readout data to propagate back as input data into the tail of the green highlights the DFF dashed line. The propagation path of the subsequent trigger to read out the tail of the orange highlights the DFF dashed line. The absolute difference between these two paths determines the shortest clock cycle time allowed.}
  \label{fig:mem-csrm}
\end{figure}

\subsubsection{Memory Access Controller}
\label{sec:mac}
The TFF-based counter sets/resets NDROs to generate trigger signals for writing and reading data into coefficient SRMs. The NTT uses a constant-geometry approach with a fixed sequence of read-write access patterns, eliminating the need for random memory access. Traditional FSMs in RSFQ logic face large feedback loop issues, making it challenging to produce control signals at the target clock frequency. Instead, TFF-based counters, arranged in a binary tree as frequency dividers, produce one-hot encoded outputs without feedback loops, only limited by the root TFF gate's maximum input rate. For $N=128$, increasing the TFF tree stages provides a 128-bit one-hot encoded output, setting/resetting NDROs on specific cycles. Regardless of $N$ (as long as it is a power of 2), the necessary outputs always come from the two leftmost TFFs, allowing pruning of other TFFs.

\begin{figure}[!t]
\centering
\includegraphics[width=3.3in]{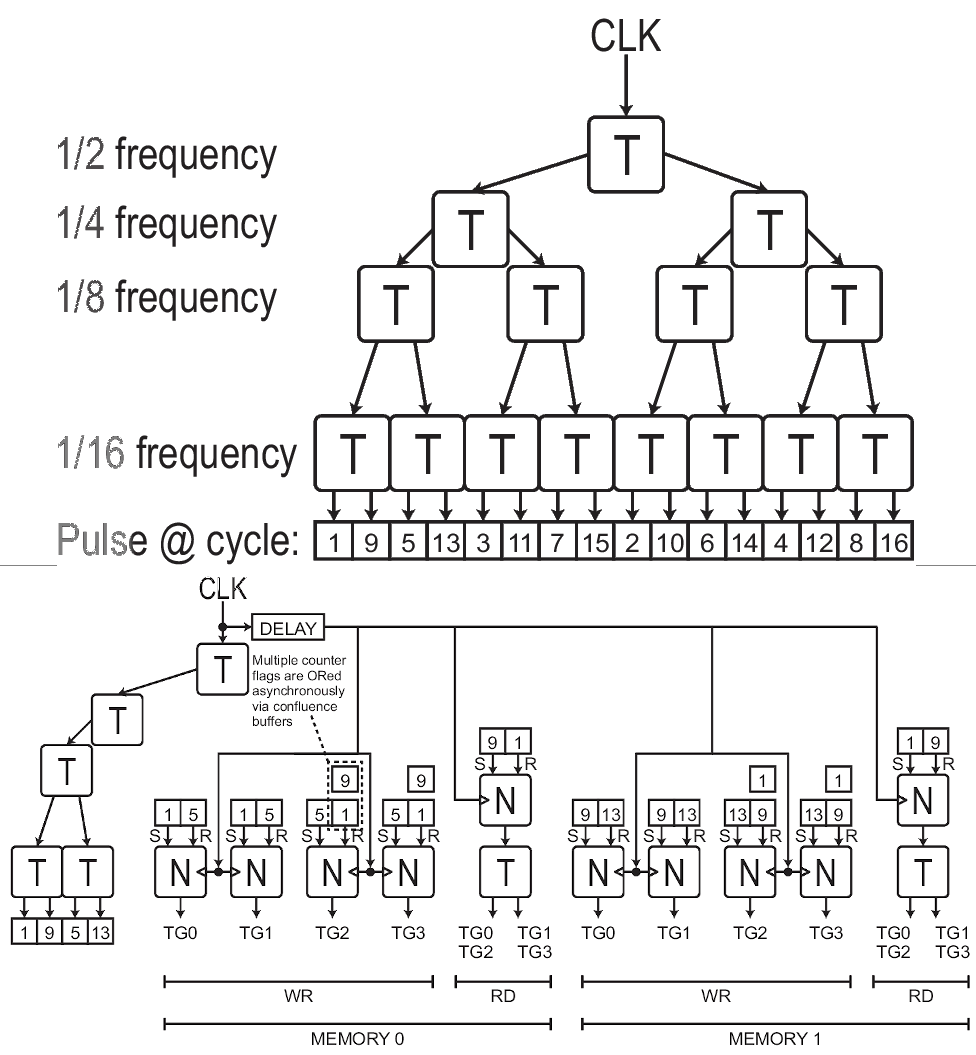}
\caption{Memory access controller (MAC) for $N=16$.}
\label{fig:mem-mac16}
\end{figure}

\subsubsection{Glue Logic}
\label{sec:glue}
We have described the design of the three major components of the memory subblock, namely: (1) the coefficient SRM, (2) the Twiddle factor CSRM, and (3) the MAC. As shown in Fig.~\ref{FIG: The NTT-128 architecture design}, several other smaller components still exist to complete the memory subblock. Miscellaneous peripheral circuits that ``glue'' the memories (SRMs and CSRMs) and MAC together. These include active delay lines for picosecond accuracy timing closure, multistage fanout trees for broadcasting signals, and asynchronous MUXes based on merger trees.

\begin{figure}[!t]
  \centering
  \includegraphics[width=3in]{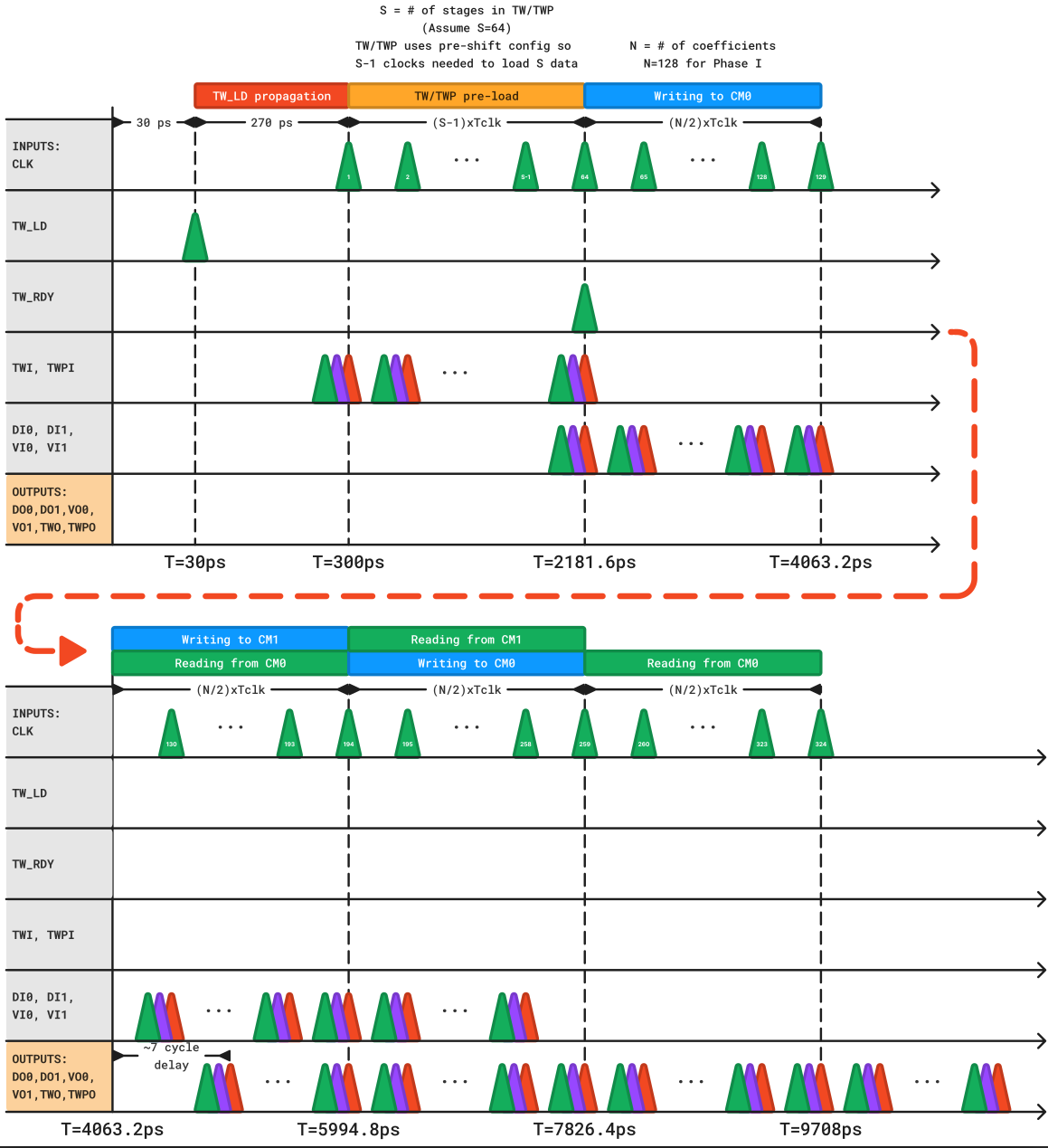}
  \caption{Example timing of signals for the memory subblock.}
  \label{fig:mem-timing}
\end{figure}

\subsubsection{Simulation and Verification}
\begin{figure}[!t]
  \centering
  \includegraphics[width=3in]{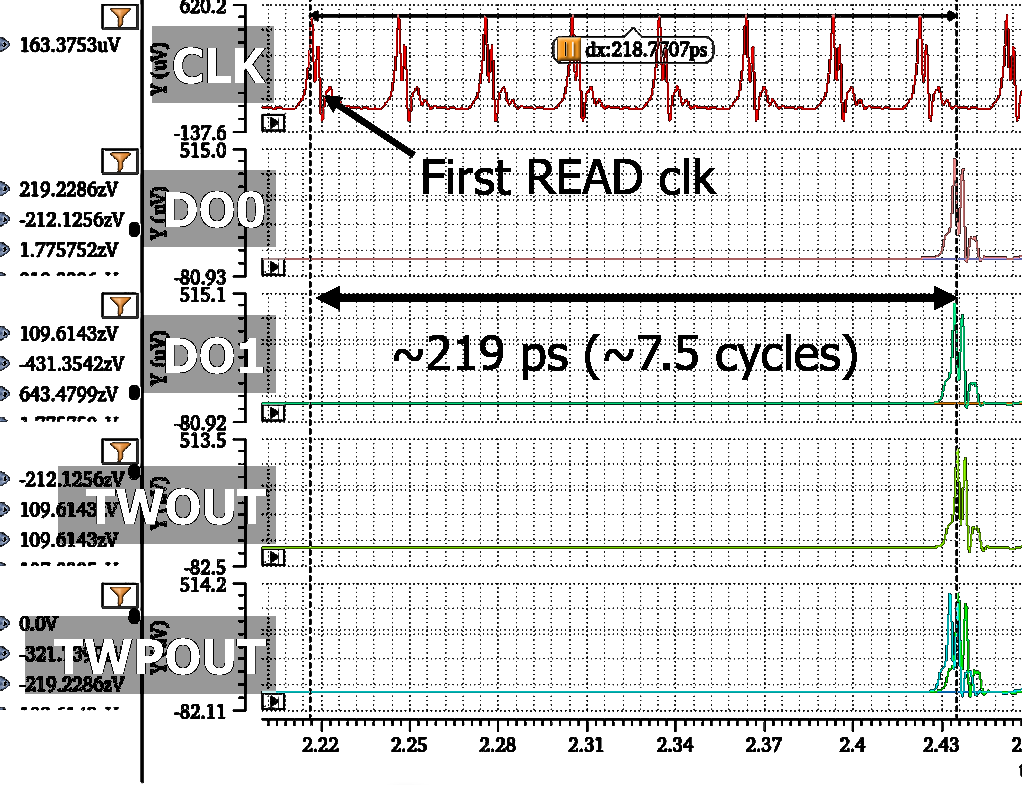}
  \caption{Simulation of the memory subblock where latency was measured.}
  \label{fig:mem-sim}
\end{figure}

The simulation of the memory subblock for each of the 7 PEs to process $N=128$ coefficients was done primarily at the analog level using JoSIM. Each memory subblock design ranged from 140000~JJs to 190000~JJs. As described in Fig.~\ref{sec:memory-infrastructure} and illustrated in Fig.~\ref{fig:mem-full-custom-flow}, an analog simulation and verification flow was established. We produced appropriate D2F JSON configuration files that generated the same input vectors used in the Python architecture model, following the timing of the signals shown in Fig.~\ref{fig:mem-timing} at 34~GHz. After the simulation, the waveforms were manually inspected to measure the overall latency and other characteristics. Fig.~\ref{fig:mem-sim} shows an example waveform used to measure the clk-to-q latency of the memory subblock. The multi-bit output traces were overlaid on each other to illustrate the small output spread achieved.

The VWrap script was configured to create a TOML verification configuration with the corresponding expected outputs obtained from the Python architecture model. All memory subblocks passed analog verification based on the I/O patterns provided by the Python model.

\section{Clocking Scheme}
SFQ circuits are highly susceptible to timing uncertainty due to process variations and environmental factors, with small logic delays that make them prone to hold violations \cite{razmkhah2024challenges}. In SFQ, a single hold violation can render a circuit inoperable, and stalling the clock does not solve the problem \cite{ISVLSI}. Inspired by traditional two-phase CMOS clocking, we extended multiphase clocking schemes to create 100\% hold-safe circuits where hold violations are resolved by adjusting clock frequencies and phase shifts. This approach uses multiple clocks with the same frequency but shifted phases, significantly reducing the need for path-balancing DFFs and enabling efficient multithreaded operation.

The clock period increases with the number of phases, introducing an area vs. throughput trade-off. For example, using ten clocks reduces the number of DFFs by more than 95\% but reduces the maximum throughput by 10x. The reduction in DFFs is achieved by assigning clock phases to gates to resolve path imbalances without DFFs. This synthesis problem is NP-hard, and previous ILP-based solutions did not scale well. We reformulated the multiphase algorithm as a linear program (LP) to improve scalability \cite{AvilesMultiPhase}.

We synthesized multiphase clocking versions of our design using this LP formulation, reducing peak performance but ensuring operability regardless of timing variations. We guarantee a satisfactory timing solution post-fabrication using multiphase clocks to drive each gate with a clock phase different from its fanouts. Our designs reduce the DFF cost by 14\% using two clock phases at 17GHz (266M NTT/Sec) or by 66\% using three clock phases at 8.5GHz (133M NTT/Sec).

\begin{figure}[!t]
\includegraphics[width=2.8in]{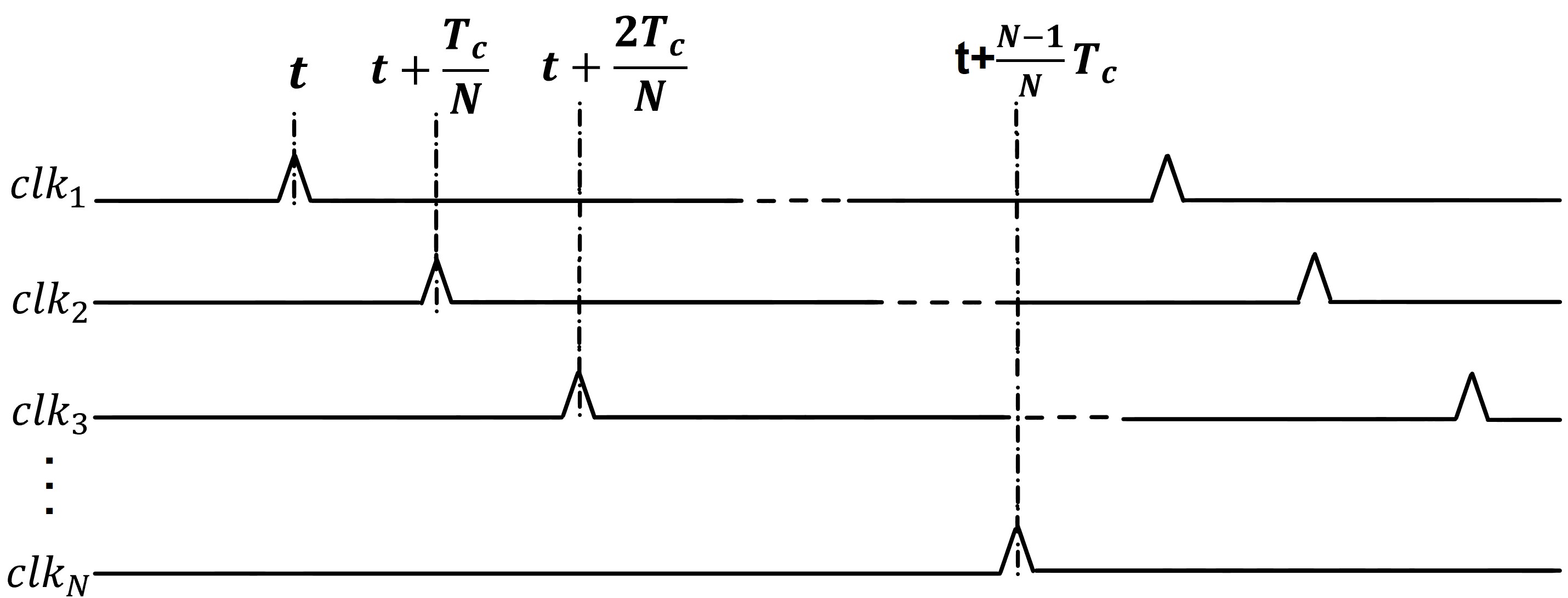}
\centering
\caption{Multi-Phase Clocks ($T_c$: cycle time)}
\label{fig:MP_clks}
\end{figure}

\begin{figure}[!t]
\includegraphics[width=2.8in]{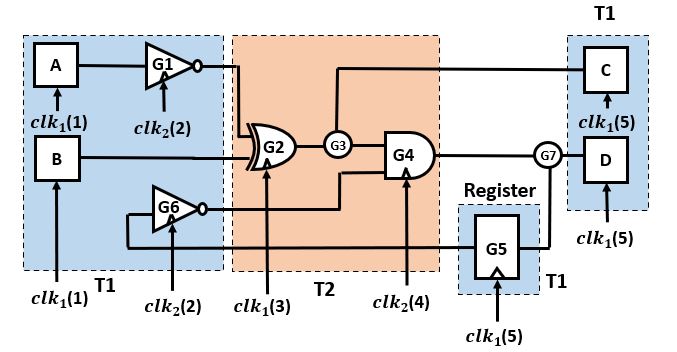}
\centering
\caption{Example Circuit Designed with Two-Phase Clocking (depth label)}
\label{fig:MPCC}
\end{figure}

\subsection{Result and System Verification}
Our verification methodology includes verification of the logic behavior of cells, considering the electrical interactions with the interconnects and cells in their fanin and fan-out configurations, followed by the characterization of the cell timing. Verilog behavioral models for the cells were also developed, and the entire design was validated using the Icarus Verilog simulation engine. Despite verifying the logic behavior for each cell independently, as described in~\cite{multicellchar}, potential logic misbehavior can still arise due to cell-to-cell and cell-to-interconnect interactions. Therefore, we have adapted the comprehensive methodology introduced in~\cite{multicellchar} to ensure that cells account for all electrical interactions within our circuits.

\subsubsection{CUS+ generation}

In~\cite{multicellchar}, the circuit under study plus (CUS+) is created for verification as follows: Initially, each logic cell in the provided cell library is chosen as the cell under study (CUS). Subsequently, numerous netlists are generated for each CUS, termed CUS+, wherein every logic cell in the library serves as the driver for each CUS input, and each logic cell in the library acts as the load on the CUS output. Additionally, our new SFQ-FHE library includes six interconnection cells, namely 1-2, 1-3, and 1-4 splitters, as well as 2-1, 3-1, and 4-1 mergers. Consequently, for each CUS+ obtained, six distinct versions are created by integrating each of the interconnection cells between each CUS and its driver(s) and load.

\begin{figure}[!t]
	\centering
	\includegraphics[width=3.5in]{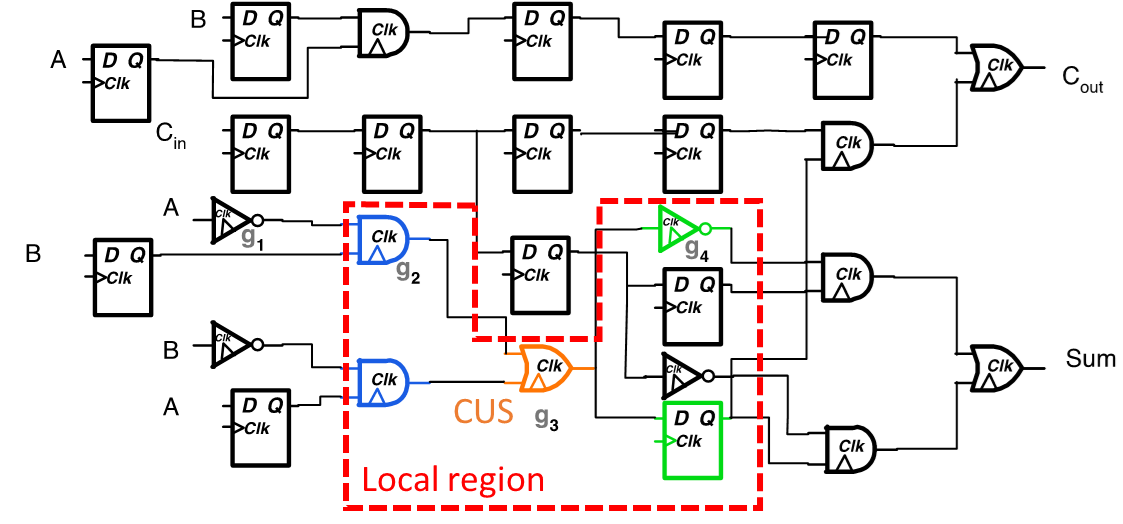}
	\caption{Sample CUS+ in a circuit under test.}
	\label{FIG:CUS+_in_situ}
\end{figure}

\paragraph{In-situ verification}
The total number of CUS+ depends on the number of cells in the library and, in our case, reaches $2\times10^{12}$ configurations. Therefore, it is impractical to verify all possible CUS+ configurations. For a feasible yet thorough method, we perform logic verification only for the CUS+ configurations in our circuit. To further accelerate verification, we skip the CUS+ that occurs only in the memory block since the entire memory block is verified in our SPICE-level simulator, JoSIM. A sample CUS+ is shown in the red block of Fig.~\ref{FIG:CUS+_in_situ}.   

\paragraph{Characterization setup}
We use JoSIM with an exhaustive set of patterns. We control the input arrival time and the clock arrival time. We use ideal voltage sources to drive the clock and the input nodes of the CUS. The output node is connected to a PTL, a PTLRX, and a resistance. 

\subsection{Module function verification}
A Verilog model is created for each cell that uses the logic function and delay values obtained from analog simulations. These models are used to verify the entire BU with Verilog simulation. Since the inputs to the BU have a special constraint, we cannot directly apply pure random patterns to the unit. Hence, we have developed a functional pattern generator and a behavioral simulator to compute the corresponding golden result generator for Verilog-based verification. In all our simulations, the waveform output by the Verilog simulator matches the corresponding golden results, establishing the functional correctness of our design as stated in equation \ref{Eq:MFV}.

\begin{equation}
\begin{aligned}
  A_{OUT} = (A_{IN} + B_{IN} * TW) \mod q \\
  B_{OUT} = (A_{IN} - B_{IN} * TW) \mod q
\end{aligned}
\label{Eq:MFV}
\end{equation}

\subsection{Architecture simulator}
To verify the proposed memory access pattern, we implemented a software simulator in Python that followed the object-oriented design method. There are two objects in detail: coefficient memory and BU. The number of pipeline stages for each module is specified. These objects are instantiated and connected, as shown in Figure \ref{FIG: The NTT-128 architecture design}. 

We first generate multiple NTT operations and store all coefficients inside the list before pushing them into the first PE. The simulation will simulate each module's inputs and outputs and store the last PE's valid outputs. Therefore, extra outputs of each module in different clock cycles are needed to help debug the hardware implementation. The golden results are then computed using the brute force method (that is, directly computing the sum of products as shown in Equation \eqref{EQ: original NTT problem}) and stored in a list. Finally, the results generated by the last processing element (PE) will be compared with the golden results. We have verified $10^5$ test cases and produced the correct results.

\section{Feasibility of FHE}

The security of FHE schemes is proven based on ring learning with errors (RLWE) theory. Given a sample $(a, b)$, where $a, b \in R_q$, the RLWE theory establishes the difficulty of distinguishing between the following two scenarios.
\begin{itemize}
    \item $ a \leftarrow R_q,  s \leftarrow R_q,  e \leftarrow \chi, b = a\times s + e$
    \item $ a \leftarrow R_q,  b \leftarrow R_q  $
\end{itemize}
where $a\leftarrow R_{q}$ denotes a uniformly random sample $a$ in $R_{q}$, $\chi$ denotes some distribution in $R_{q}$, and $\times$ represents the polynomial multiplication.

The original CKKS scheme \cite{b_CKKS} is a leveled FHE scheme, where each level corresponds to a specific computation, such as a homomorphic multiplication. The ciphertext modulus at the first level is expressed as $q_l = p^l \cdot q_0$, with $0 \leq l \leq L$, $p \approx q_0$. Typically, \textbf{p} is a 30-bit prime number. The multiplication level $(l)$ represents the maximum number of homomorphic multiplications performed before the data is corrupted. The result after each level of computation is divided by a scaling factor \textbf{p} to ensure the scheme's security; errors are intentionally embedded in the least significant bits of coefficients within the ciphertexts. The ciphertext modulus decreases after each computation level to reduce these errors.

The size of the ring, denoted by the degree of the polynomial \textbf{N}, and the bitwidth of the modulus collectively contribute to the security level captured by parameter $\lambda$. A $\lambda$-bit security system requires the attacker to perform $2^{\lambda}$ operations to break it. In the original CKKS scheme, the relation between \textbf{N}, $\log_2(q_L)$, and $\lambda$ is expressed as,
\begin{equation}
    N \geq \frac{\lambda + 110}{7.2} \log(P\cdot q_L)
\end{equation}
where \textbf{P} is a special integer generated during the setup phase of the system and is used for homomorphic multiplications only. In addition, the value of \textbf{P} is approximately equal to $q_L = p^{L} q_0$.
Reference \cite{b_CKKS} sets $N= 2^{13}$, $\log(q_L) = 155$, and $\log(p) = 30$ to evaluate $x^{16}$ using $\log_2(16)=4$ multiplications and to maintain an 80-bits security level. 

The original CKKS scheme requires a large bitwidth for the ciphertext modulus to ensure security, increasing the hardware cost. Reference \cite{b_CKKS19} presented the residue number system (RNS) version of the CKKS scheme, where a sizeable large bitwidth modulus is the product of multiple small-bitwidth moduli. Let $\mathcal{B} = \{p_0, p_1, p_2, \cdots, p_{k-1}\}$ denote the basis and $P = \prod_{i=0}^{k-1} p_i$. The symbol $[\cdot]_{\mathcal{B}}$ represents the transformation from $\mathbb{Z}_P$ to $\prod_{i=0}^{k-1} \mathbb{Z}_{p_i}$ where $[a]_{\mathcal{B}} = ([a]_{p_i})$ with ${0\leq i \leq k-1}$. The ring isomorphism over the integers can be naturally extended to a ring isomorphism $[\cdot]_\mathcal{B}$ (i.e., the map: $R_P \rightarrow R_{p_0} \times R_{p_1} \times \cdots R_{p_{k-1}} $) by applying it coefficient-wise over the cyclotomic rings. Consequently, computations can be conducted separately based on these small moduli, leading to a more compact and efficient datapath. Meanwhile, the original ciphertext under the large modulus can be treated as multiple small ciphertexts under the small moduli. 

\section{Large Scale Implementation}
To perform a practical-size NTT operation ($N = 2^{14}$) using our proposed hardware, we need to perform two stages of computations, each stage performing 128 SCE-NTT (128-point NTT) for a total of 256 SCE-NTTs, utilizing the divide-and-conquer property. In other words, by employing the proposed NTT-128 architecture, we can execute a $2^{14}$-point NTT operation in $256 \times 64 = 16,384$ clock cycles with appropriate scheduling. Using ideal conditions for the SCE circuit, the clock frequency is limited by the NTT structure. Considering that the NTT-128 operates at 34~GHz, the latency will be 482~ns. In comparison, reference \cite{b_HEAX} requires approximately 23,894~ns to compute one $2^{14}$-point NTT operation, operating at 300~MHz.

Furthermore, we can leverage the high performance of the proposed design for NTT of any size with the help of additional computation units. For instance, to perform a $2^{15}$-point NTT operation, we may divide it into two sub-problems and then apply the proposed NTT-128 to execute $2^{14}$-point NTT operations. The butterfly operations of the merge step will be performed in the application software. Therefore, the proposed NTT-128 architecture can handle different configurations. For a practical FHE system with a ring size of $2^{14}$ and 60-bit datapath bitwidth, we can apply our proposed SCE NTT-128 design after extending it to the $60$-bit datapath. Based on the divide-and-conquer method, a $2^{14}$-point NTT operation can be divided into $2^7$ subproblems, where each subproblem is a $2^7$-point NTT. 

\begin{figure}[!t]
	\centering
	\includegraphics[width=2.2in]{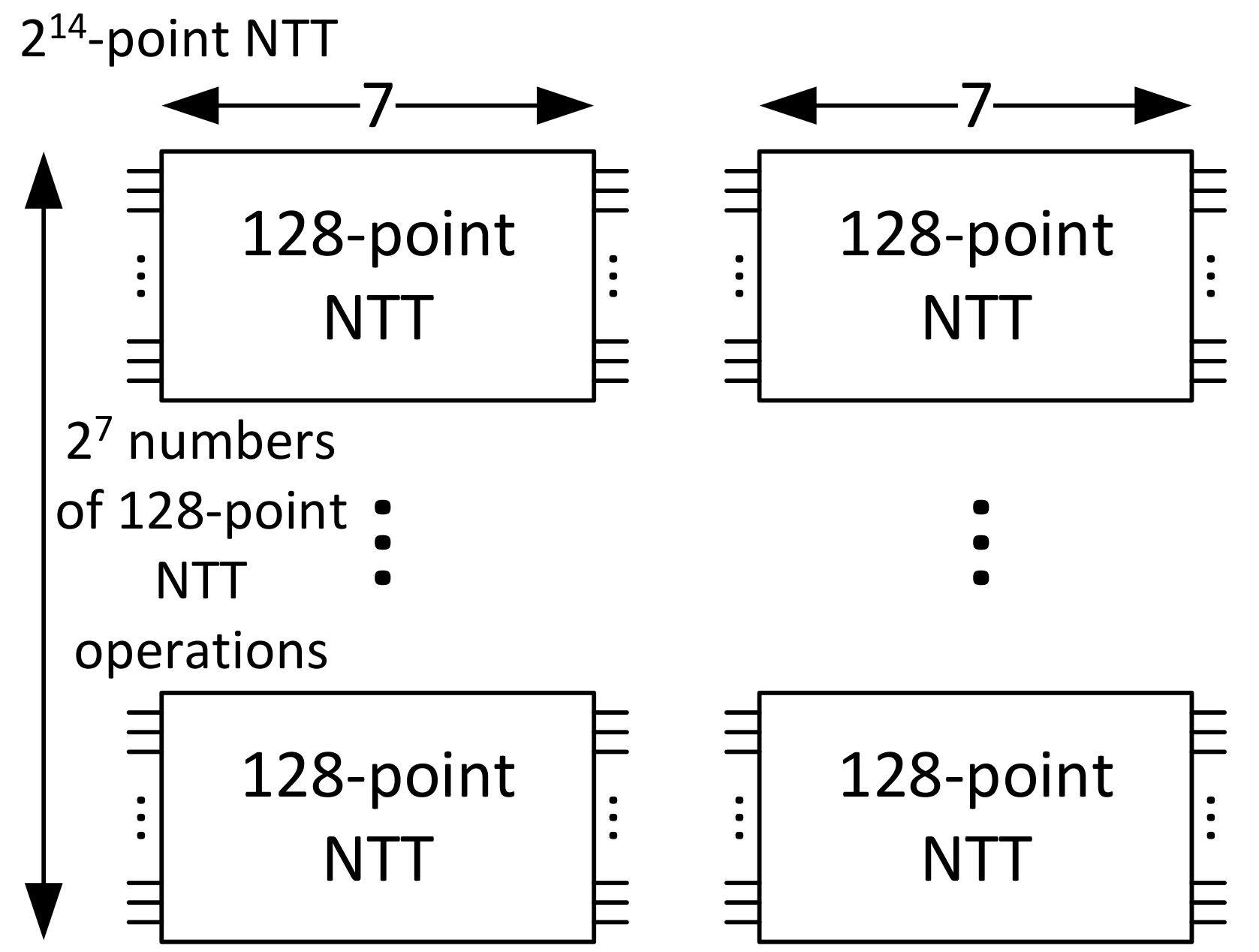}
	\caption{The $2^{14}$-point NTT operation performed by $128$-point NTT operations}
	\label{FIG: The 2^14-point NTT operation performed by 128-point NTT operations}
\end{figure}

Meanwhile, the merging of these subproblems is effectively handled by the NTT-128 design. A $2^{14}$-point NTT operation is divided into $14$ stages, each involving $2^{13}$ number of butterfly operations. The NTT-128 design can be employed to execute these butterfly operations across two stages, where each stage comprises $2^7$ instances of the SCE NTT-128. This operational flow is illustrated in Figure \ref{FIG: The 2^14-point NTT operation performed by 128-point NTT operations}.

Due to the dependency between the first and second stages of the NTT-128 operations, a waiting period of approximately 200 clock cycles is necessary to ensure that all data is flushed from the pipeline registers before one can start the second computation stage. Consequently, the $2^{14}$-point NTT operation requires approximately $(128 \times 64 \times 2 + 400 \approx 16,784)$ clock cycles. We may instantiate $K$ NTT-128 designs operating in parallel to reduce the cycle counts. In this situation, the total clock cycle count will be approximately $((128 \times 64/K) \times 2 + 400)$ while ignoring clock cycles for the reorder operation to generate correct inputs. 
However, it is important to consider that increasing the number of SCE NTT-128 cores also increases the required memory bandwidth and complexity of the multiplexer design, leading to a larger area. This approach similarly applies to the inverse NTT operations, where using the inverse NTT-128 module is expected to show comparable performance and hardware requirements as the NTT-128.

For a homomorphic multiplication, two architectures were investigated. First is reusing the same NTT units to perform all NTT operations of homomorphic multiplication operations sequentially \cite{b_Trebuchet}. This approach is characterized by a complex data flow and intricate data dependencies, necessitating the deployment of complex finite-state machines and network structures to provide the correct data. In contrast, the architecture demonstrated in \cite{b_HEAX}, pioneered by the Microsoft SEAL team, advocates for instantiating multiple distinct arithmetic units. Each unit is dedicated to a specific step within the homomorphic multiplication process. Therefore, the data rerouting network is simplified, and the throughput of homomorphic multiplication is increased.

We estimated the arithmetic cost of homomorphic multiplication using the proposed design in SCE circuits for different operations (pointwise multiplication and key switch operation) while ignoring the intermediate memory, data reroute network, and finite state machine. 

\begin{figure}[!t]
	\centering
	\includegraphics[width=3.5in]{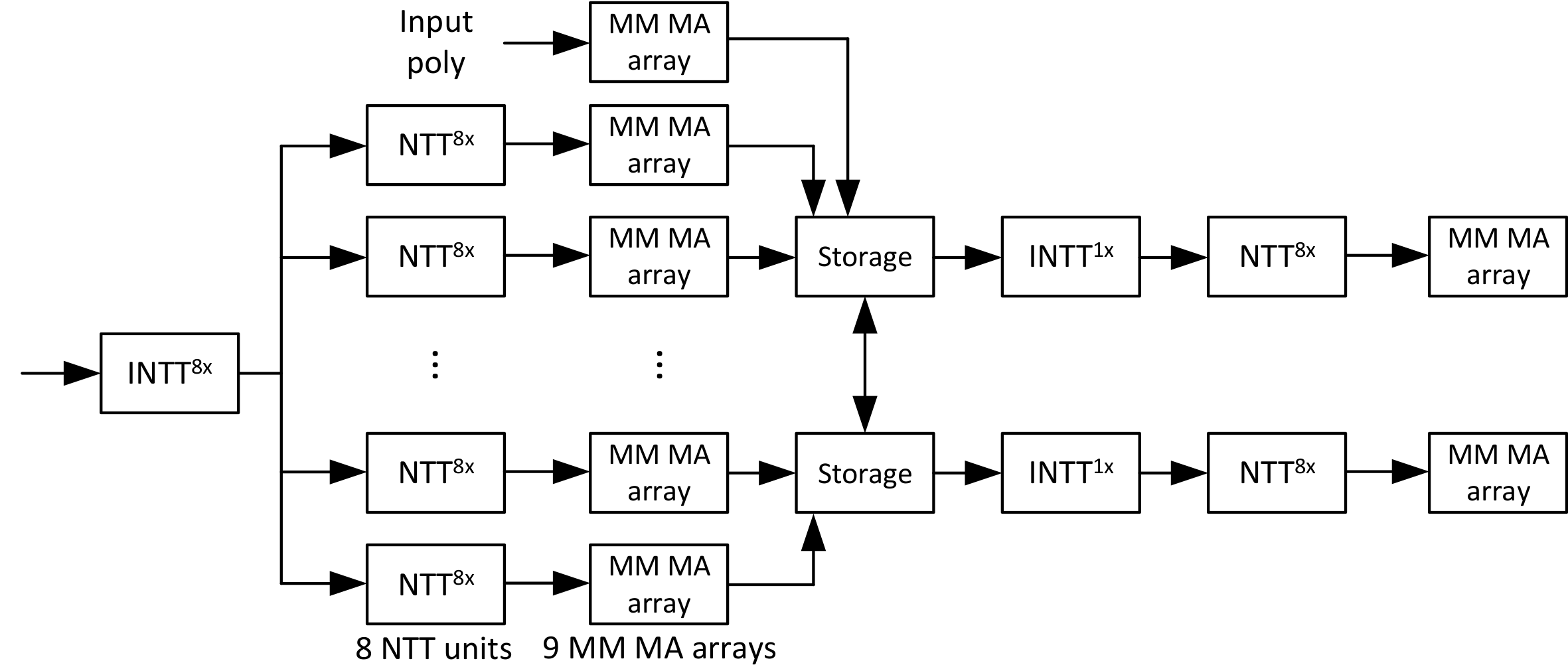}
	\caption{The architecture of key switch operations}
	\label{FIG: The architecture of key switch operations}
\end{figure}

Figure \ref{FIG: The architecture of key switch operations} shows a pipelined architecture design that can perform the key switch operation for $N = 2^{14}$ and $L+1 = 8$. All modules are pipelined, and each module is responsible for one operation inside the outer loop of the key switch operation. Computations in different outer loop iterations can be sequentially assigned to the proposed architecture, one after the other. 

The $\text{NTT}^{8\times}$ represents that there are eight NTT-128 designs instantiated in the NTT unit for performing $2^{14}$-point NTT operation. 
The modular multiplication (MM) and modular addition (MA) array comprises multiple modular multipliers and modular adders to perform "Dyadic Mod" operations. 
Based on this architecture, we can derive the timing diagram of the key switch operation.

As depicted in Figures \ref{FIG: The architecture of key switch operations}, our design integrates eight INTT-128 units within the INTT framework, achieving an approximate latency of $2,600$ cycles. Next, we enable performing $L+1 = 8$ NTT operations within an outer iteration using eight parallel NTT units. Each unit comprises eight NTT-128 modules, aligning its latency with that of the INTT unit.
Moreover, we place two modular multipliers at the primary input stage of each NTT-128 unit. These multipliers execute modular multiplications before the NTT operation. 

There are $2 \times 9$ pointwise multiplications to perform "Dyadic Mod" operations (ignoring the add operations since the hardware area of modular addition is relatively small.) Consequently, we employ nine MM/MA arrays. Each array is equipped with 14 modular multipliers and adders, capable of performing $2 \times 2^{14}$ modular operations. This setup achieves a processing time of approximately $2,400$ cycles.
The topmost MM/MA array is designated for the 'Dyadic Mod' operation, bypassing the NTT process.

After all these intermediate results are stored in memory, we perform the two RNS floor operations via two parallel datapaths. Given that only one NTT-128 design is instantiated to perform the INTT operation, approximately $17,000$ clock cycles are required, as previously discussed. The final stage involves the MM/MA array executing the MS operation. This array comprises roughly eight modular multipliers and 16 modular adders, culminating in a latency of around $2,600$ cycles. 

In summary, our proposed architecture, requiring some 90 NTT-128 modules, 302 modular multipliers, and 150 modular adders (under 60-bit datapath), ensures high throughput, enabling the execution of a key switch operation within $20,800$ clock cycles $(2600 \times 8)$. Assuming an operational frequency of 34 GHz for the SFQ circuit, the key switch operational throughput is projected to be $(10^9 / (0.0294 \times 20800)) = 1,634,614$ per second. This throughput significantly surpasses the reference \cite{b_HEAX}, which achieved an approximate throughput of $2,616$ key switch operations per second.

\section{Conclusion}
We have successfully designed and verified a 128-point 32-bit NTT accelerator using superconductor logic that has achieved outstanding results, surpassing CMOS designs by over 100 times, reaching 531M NTT/sec at 34GHz. A new RSFQ cell library featuring 50 cells, including complex multistage compound cells, was developed to achieve this goal. We have ensured the functional correctness and timing of the cells through validation using JOSIM and Verilog. Key components like shift register memory blocks and modular arithmetic units have been designed, validated, and optimized for performance. Our approach to dual-phase clocking has kept the circuits' high performance while reducing path balancing overhead by 14\% and ensuring robustness for hold violations. We have also demonstrated a large-scale implementation architecture for $2^{14}$ points NTT and INTT using the proposed hardware, achieving 1.63 MOP/s, surpassing all other designs. While the current SCE fabrication process does not support such complex circuits, the rapid developments in fabrication will allow the utilization of these circuits in the near future. These achievements underscore the potential of superconducting electronics in high-performance computing and secure communication for the post-quantum era.

\section*{Acknowledgments}
This work is supported by the Defense Advanced Research Projects Agency (DARPA) under the SCE-NTT: Hardware Accelerator for Homomorphic Computing Utilizing Superconductor Electronics project.

\newpage

\vfill

\end{document}